\documentclass[aip,twocolumn,reprint,nofootinbib]{revtex4-1}
\usepackage{graphicx}
\usepackage{amsmath}
\usepackage{amssymb}
\usepackage{comment}
\usepackage{tikz}
\usepackage{lipsum}
\usepackage{subcaption}
\usepackage{newfloat}
\usepackage{algorithmic}
\usepackage{algorithm}
\usepackage{color}

\newcommand{\ket}[1]{| #1 \rangle}
\newcommand{\bra}[1]{\langle #1 |}
\newcommand{\braket}[1]{\langle #1 \rangle}
\newcommand{\hc}{\text{H.c.}}

\newcommand{\eq}[1]{\begin{align}#1\end{align}}
\newcommand{\nn}{\nonumber}
  
\newcommand{\uu}{\mathcal{\hat U}}
\newcommand{\inn}{\text{in}}
\newcommand{\out}{\text{final}}
\newcommand{\seq}[1]{\begin{subequations}#1\end{subequations}}
\newcommand{\sseq}[1]{\seq{\eq{#1}}}
\newcommand{\vac}{\bm{0}}

\newcommand{\diag}{\text{diag}}

\newcommand{\D}{\mathcal{\hat D}}
\newcommand{\s}{\mathcal{\hat S}}
\newcommand{\id}{\mathbb{\hat I}}
\newcommand{\pp}{\left(\bigotimes_{j=1}^{2\ell} \a_j^{p_j} \right)}
\newcommand{\ppp}{\left(\bigotimes_{l=1}^P \b_j \right)}
\usepackage{bm}
\newcommand{\lhaf}{\text{lhaf}}
\newcommand{\haf}{\text{haf}}
\renewcommand{\a}{\hat a}
\newcommand{\ad}{\hat a^\dagger}
\renewcommand{\b}{\hat b}
\newcommand{\bd}{\hat b^\dagger}
\newcommand{\R}{\bm{\hat R}}
\renewcommand{\P}{\bm{\hat P}}
\newcommand{\Rn}{\bm{\hat R}}
\newcommand{\Pn}{\bm{\hat P}}

\newcommand{\ea}{\emph{et al.}}
\newcommand{\ta}{\bm{\tilde \alpha}}
\newcommand{\tu}{\bm{\tilde U}}
\newcommand{\tl}{\bm{\tilde \Lambda}}
\newcommand{\rr}{\lambda}
\newcommand{\bp}{B^{(\bm{p})}}
\newcommand{\zp}{\zeta^{(\bm{p})}}
\begin{document}

\title{
	Franck-Condon factors by 
	counting perfect matchings of graphs with loops}
\author{Nicol\'as Quesada}
\email{nicolas@xanadu.ai}
\affiliation{Xanadu, 372 Richmond Street W, Toronto, Ontario M5V 1X6, Canada}

\begin{abstract}
We show that the Franck-Condon Factor (FCF) associated to a transition between initial and final vibrational states in two different potential energy surfaces, having $N$ and $M$ vibrational quanta, respectively, is equivalent to calculating the number of perfect matchings of a weighted graph with loops that has $P = N+M$ vertices. This last quantity is the loop hafnian of the (symmetric) adjacency matrix of the graph which can be calculated in $O(P^3 2^{P/2})$ steps. In the limit of small numbers of vibrational quanta per normal mode our loop hafnian formula significantly improves the speed at which FCFs can be calculated. Our results more generally apply to the calculation of the matrix elements of a bosonic Gaussian unitary between two multimode Fock states having $N$ and $M$ photons in total and provide a useful link between certain calculations of quantum chemistry, quantum optics and graph theory.
\end{abstract}

\maketitle
\section{Introduction}

Vibronic spectroscopy is a fundamental tool for probing molecules\cite{herzberg1945infrared,herzberg1988spectra}. By studying the energy distribution of the light emitted while molecules change their vibrational and electronic states, a great deal of structural molecular information can be obtained. The inverse problem, to design a molecule by tailoring the frequencies at which it emits or absorbs light, is of paramount importance. Having efficient tools for this inverse problem would allow the fast assessment of candidates for dyes in biological and industrial settings and for solar cells in the energy industry\cite{gross2000improving,hachmann2011harvard}.

Effective time-independent approaches have been extensively used to simulate vibrationally resolved electronic absorption/emission spectra of molecules\cite{biczysko2012time,lami2011time} .  In this framework, line intensities are given by the matrix elements of the electronic transition dipole moment calculated over the initial and final vibrational wave functions. Within the harmonic approximation, the vibrational wave functions are the eigenstates of the Hamiltonian describing the uncoupled harmonic oscillations of the nuclei around their equilibrium positions in the initial and final potential energy surfaces. If one assumes no dependence of the electronic transition dipole with respect to the changes in the molecular geometry during the transition, the so-called Franck-Condon (FC) approximation, the vibrational structure of the optical bands is determined by the overlap integral of vibrational wave functions, known as the FC factors\cite{ruhoff1994recursion}. In order to use a common coordinate system to evaluate the latter integrals, Duschinsky\cite{duschinsky1937importance} proposed a linear transformation to account for the rotation and displacement of the normal modes between the initial and final states molecular structures.

Recently, Huh \ea\cite{huh2015boson} \ showed an elegant mapping between  FC transitions and the scattering of quantum states of photons going through an interferometer and measured using photon number resolving detectors (PNRD). They showed that FCFs are analogous to the photon-number probability amplitudes of squeezed-displaced states after going through a linear optical network.
A similar problem, $N$ single photons scattering through an interferometer and being probed by PNRD, so-called Boson Sampling\cite{aaronson2011computational}, has provided a significant momentum for the development of photonic technologies and for the study of combinatorial sampling problems. These developments  were motivated by Aaronson and Arkhipov\cite{aaronson2011computational}, who showed that a classical computer requires a time that is exponential in the number of photons to generate the samples that the quantum experiment generates by construction. This was one of the first provable computational tasks in which a rudimentary, non-universal quantum computer can exhibit an exponential speedup over classical computers\cite{harrow2017quantum}.

An important part of the arguments developed by Aaronson and Arkhipov is that the probability amplitudes appearing in Boson Sampling are proportional to a matrix function known as the permanent, which counts the number of perfect matchings of an undirected bipartite graph, and is in the $\#P$ Complete complexity class\cite{valiant1979complexity}.

In analogy with Boson Sampling, one might wonder if there is a matrix function, related to graphs, that is proportional to the FCFs of a given vibronic molecular transition. An important step in answering this question was taken by Hamilton \ea \cite{hamilton2017gaussian} and Kruse \ea \cite{kruse2018detailed} Using phase-space methods, they showed that, for the case of squeezed states (with no displacement) going into a linear optical network and probed with PNRD, the probability (not the amplitude) of given Fock number detection event is proportional to another matrix function called the hafnian. The hafnian counts the number of perfect matchings of an undirected graph without loops. Although this is an important advance, it does not solve the question that was posed at the beginning of this paragraph; for FCFs, the displacements are critical since they describe how the equilibrium geometry is changed as the light is absorbed or emitted. Indeed, the displacements are zero only if all equilibrium positions before and after the light is absorbed are the same. Also, the phase-space methods developed by Hamilton \ea \ and Kruse \ea \ can only give probabilities, not probability amplitudes.

In this paper, we analyze the complexity of FCFs by mapping their evaluation to the calculation of matrix elements of a certain Gaussian quantum unitary operator. Secondly, using methods that do not require any knowledge of the quantum phase-space, we show that the probability amplitudes of these quantum optical unitaries are equal to the loop hafnian matrix function of a certain matrix; this function counts the number of perfect matchings of an undirected graph \emph{with loops}. Thus our methods generalize the results of Hamilton \ea by allowing for the inclusion of displacements and also by giving directly probability amplitudes.

Analytical forms of the matrix elements of Gaussian unitaries have been derived by Dodonov \ea \cite{dodonov1994multidimensional} and Kok \ea\cite{kok2001multi}. As noted by Philips \ea\cite{phillips2018certification}, the calculation of these matrix elements ``can be done using multidimensional Hermite polynomials but involves rather complicated computations.'' 
Our method reduces these calculations to simple manipulations of 
adjacency matrices of graphs and the use of the loop hafnian function.
The reduction of the problem of FCFs to loop hafnians also allows us to provide a tight bound on how fast FCFs can be calculated. Indeed, we show that for a molecule in which $N$ phonons of the ground electronic surface are overlapped with $M$ phonons of the excited electronic surface, the complexity of their FCF will scale like  $O(P^3 2^{P/2})$ with $P=N+M$. For the zero temperature case, in which initially there are zero bosons being scattered, $N=0$, the time scales like $O(M^3 2^{M/2})$. These results provide what we believe are some of the fastest methods to calculate FCFs in the harmonic approximation. 

In order to make clear how graph theory and quantum optics can be used to tackle the calculation of the vibronic spectrum of molecules, we have structured the manuscript as follows:
in the next three sections, we introduce basic notation relating to graph theory and the hafnian and loop hafnian functions (Sec.~\ref{sec:graphs}), creation and annihilation operators and commonly used Gaussian bosonic operations (Sec.~\ref{sec:HO}), and finally we use this notation to set up the problem of calculating FCFs in the harmonic approximation (Sec.~\ref{sec:FCFs}). In Sec.~\ref{sec:amplitudes}, we carry out the proof of the main result: relating FCFs to loop hafnians. This is accomplished in three steps. First, by a judicious use of operator-disentangling theorems, we rewrite the action of the Gaussian operators in terms of bosonic creation operators only. Secondly, we relabel modes in which multiple bosons are measured into new ancillary modes into which only single bosons are measured. Thirdly, using the multinomial theorem, we contract the resulting expression to show that, indeed, the amplitude is a loop hafnian.
At the end of this section we provide a summary of the results which should  allow the straightforward calculation of any FCF using the \texttt{hafnian} library\cite{hafnian}.
In Sec.~\ref{sec:discussion}, we present a discussion of the algorithmic complexity of calculating FCFs and matrix elements of Gaussian unitaries.
Perhaps more importantly, we argue that the results presented in this manuscript provide a useful link and common language which should allow different scientific communities to ``import'' the algorithms from other communities which were developed for what, at first glance, might be disparate computational problems: counting perfect matchings of graphs with loops and calculating Franck-Condon factors.

A short comment on notation: we use boldface letters to denote quantities with indices, such as matrices or vectors. We use upper-case letters for matrices and lower-case letters for vectors, except for $\bm{R}$ and $\bm{P}$, denoting the position and momenta of the nuclei.
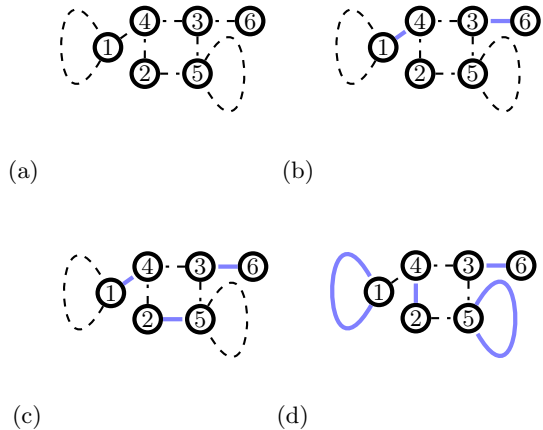
\begin{figure}[!t]
	\begin{subfigure}[b]{0.5\textwidth}
(a)
\begin{tikzpicture}[scale=.7, shorten >=1pt, auto, node distance=1cm, ultra thick]
    \tikzstyle{node_style} = [circle,draw=black, inner sep=0pt, minimum size=10pt]
    \tikzstyle{edge_style} = [-,draw=black, line width=2, thick, dashed]
    \tikzstyle{edge_styleg} = [-,draw=blue!50, line width=2, ultra thick]

    \node[node_style] (v0) at (-0.7,0.5) {1};
    \node[node_style] (v1) at (0,0) {2};
    \node[node_style] (v2) at (1,1) {3};
    \node[node_style] (v3) at (0,1) {4};
    \node[node_style] (v4) at (1,0) {5};
    \node[node_style] (v5) at (2,1) {6};
    \draw[edge_style]  (v0) edge (v3);
    \draw[edge_style]  (v1) edge (v3);
    \draw[edge_style]  (v1) edge (v4);
    \draw[edge_style]  (v2) edge (v3);
    \draw[edge_style]  (v2) edge (v4);
    \draw[edge_style]  (v5) edge (v2);

    \draw[edge_style]  (v4) to [loop right, in=-60,out=60,looseness=10] (v4);
    \draw[edge_style]  (v0) to [loop left, in=120,out=-120,looseness=10] (v0);
    
\end{tikzpicture}
\
(b)
\begin{tikzpicture}[scale=.7, shorten >=1pt, auto, node distance=1cm, ultra thick]
\tikzstyle{node_style} = [circle,draw=black, inner sep=0pt, minimum size=10pt]
\tikzstyle{edge_style} = [-,draw=black, line width=2, thick, dashed]
\tikzstyle{edge_styleg} = [-,draw=blue!50, line width=2, ultra thick]

\node[node_style] (v0) at (-0.7,0.5) {1};
\node[node_style] (v1) at (0,0) {2};
\node[node_style] (v2) at (1,1) {3};
\node[node_style] (v3) at (0,1) {4};
\node[node_style] (v4) at (1,0) {5};
\node[node_style] (v5) at (2,1) {6};
\draw[edge_styleg]  (v0) edge (v3);
\draw[edge_style]  (v1) edge (v3);
\draw[edge_style]  (v1) edge (v4);
\draw[edge_style]  (v2) edge (v3);
\draw[edge_style]  (v2) edge (v4);
\draw[edge_styleg]  (v5) edge (v2);

\draw[edge_style]  (v4) to [loop right, in=-60,out=60,looseness=10] (v4);
    \draw[edge_style]  (v0) to [loop left, in=120,out=-120,looseness=10] (v0);
\end{tikzpicture}
\\
(c)
\begin{tikzpicture}[scale=.7, shorten >=1pt, auto, node distance=1cm, ultra thick]
\tikzstyle{node_style} = [circle,draw=black, inner sep=0pt, minimum size=10pt]
\tikzstyle{edge_style} = [-,draw=black, line width=2, thick, dashed]
\tikzstyle{edge_styleg} = [-,draw=blue!50, line width=2, ultra thick]

\node[node_style] (v0) at (-0.7,0.5) {1};
\node[node_style] (v1) at (0,0) {2};
\node[node_style] (v2) at (1,1) {3};
\node[node_style] (v3) at (0,1) {4};
\node[node_style] (v4) at (1,0) {5};
\node[node_style] (v5) at (2,1) {6};
\draw[edge_styleg]  (v0) edge (v3);
\draw[edge_style]  (v1) edge (v3);
\draw[edge_styleg]  (v1) edge (v4);
\draw[edge_style]  (v2) edge (v3);
\draw[edge_style]  (v2) edge (v4);
\draw[edge_styleg]  (v5) edge (v2);

\draw[edge_style]  (v4) to [loop right, in=-60,out=60,looseness=10] (v4);
\draw[edge_style]  (v0) to [loop left, in=120,out=-120,looseness=10] (v0);
\end{tikzpicture}
(d)
\begin{tikzpicture}[scale=.7, shorten >=1pt, auto, node distance=1cm, ultra thick]
\tikzstyle{node_style} = [circle,draw=black, inner sep=0pt, minimum size=10pt]
\tikzstyle{edge_style} = [-,draw=black, line width=2, thick, dashed]
\tikzstyle{edge_styleg} = [-,draw=blue!50, line width=2, ultra thick]

\node[node_style] (v0) at (-0.7,0.5) {1};
\node[node_style] (v1) at (0,0) {2};
\node[node_style] (v2) at (1,1) {3};
\node[node_style] (v3) at (0,1) {4};
\node[node_style] (v4) at (1,0) {5};
\node[node_style] (v5) at (2,1) {6};
\draw[edge_style]  (v0) edge (v3);
\draw[edge_styleg]  (v1) edge (v3);
\draw[edge_style]  (v1) edge (v4);
\draw[edge_style]  (v2) edge (v3);
\draw[edge_style]  (v2) edge (v4);
\draw[edge_styleg]  (v5) edge (v2);

\draw[edge_styleg]  (v4) to [loop right, in=-60,out=60,looseness=10] (v4);
\draw[edge_styleg]  (v0) to [loop left, in=120,out=-120,looseness=10] (v0);
\end{tikzpicture}
\end{subfigure}
	\caption{\label{fig:graphs} (a) A graph with 6 vertices and 8 edges, two of which are loops. (b) A matching (1,4)(3,6) represented with thick blue lines; the edges in the matching do not share any vertex.  (c) A perfect matching (1,4)(2,5)(3,6) with no loops (d) A perfect matching with two loops (1,1)(2,4)(3,6)(5,5). Note that the matching in (b) is not perfect since there are unpaired vertices (2,5). }
\end{figure}

\section{Graphs, hafnians and loop hafnians}\label{sec:graphs}

In this section we introduce some basic terminology of graph theory to define the hafnian (haf) and loop hafnian (lhaf) functions.

A graph is an ordered pair $G=(V,E)$ where $V$ is the set of vertices, and $E$ is the set of edges linking the vertices of the graph, i.e., if $e \in  E$ then $e=(i,j)$ where $i,j \in  V$.
In this manuscript we will consider graphs with loops, thus we allow for edges $e = (i,i) \in  E$ connecting a given vertex to itself. A 6 vertex graph is shown in Fig.~\ref{fig:graphs}(a). The vertices are labelled $V = \{1,2,3,4,5,6 \}$ and the edges are $E=\{(1,1),(1,4),(2,4),(2,5),(3,4),(3,5),(3,6),(5,5) \}$.

A matching $M$ is a subset of the edges in which no two edges share a vertex. An example of matching, $M=(1,4)(3,6)$ is shown in Fig.~\ref{fig:graphs}(b). A perfect matching is a matching which matches all the vertices of the graph, such as for example $M=(1,4)(2,5)(3,6)$ or $M=(1,1)(2,4)(3,6)(5,5)$ in Fig.~\ref{fig:graphs}(c) and Fig.~\ref{fig:graphs}(d) respectively.

A complete graph is a graph where every vertex is connected to every other vertex. 
For loopless graphs having $n$ vertices, the number of perfect matchings is\cite{barvinok2016combinatorics}
\eq{
|\text{PMP}(n)|=(n-1)!! = 1 \times 3 \times  \ldots \times (n-1).
}
where we use $\text{PMP}(n)$ to indicate the set of perfect matchings of a complete graph with $n$ vertices, and the notation $|V|$ to indicate the number of elements in the set $V$. Note that this number is nonzero only for even $n$, since for odd $n$ there will always be one unmatched vertex.
If we consider a graph with vertices labelled $V=\{1,2,3,4 \}$ then the set of perfect matchings is 
\eq{
\text{PMP}(4)=\{(1,2)(3,4) , (1,3)(2,4),(1,4)(2,3) \}.	
}
For the case of a complete graph with loops, the number of perfect matchings of a graph with an even number of vertices is given by\cite{bjorklund2018faster}
\eq{
|\text{SPM}(n)| = T\left(n,-\tfrac{1}{2},\tfrac{1}{\sqrt{2}}\right),
}
where $\text{SPM}(n)$ is the set of perfect matchings of a complete graph with loops and $T(a,b,c)$ is the Toronto function\cite{abramowitz1964handbook}.
If we consider a graph with vertices labelled $V=\{1,2,3,4 \}$ then the set of perfect matchings including loops is 
\eq{
	\text{SPM}(4)=&\{(1,2)(3,4) , (1,3)(2,4),(1,4)(2,3),\nonumber \\
	&\ \ (1,2)(3,3)(4,4),(1,1)(2,2)(3,4) ,  \nonumber \\
	&\ \ (1,3)(2,2)(4,4), (1,1)(3,3)(2,4),  \\
	&\ \ (1,4)(2,2)(3,3), (1,1)(4,4)(2,3), \nonumber \\
	&\ \ (1,1)(2,2)(3,3)(4,4) \} \nonumber .
}

\begin{figure}[!t]
	\input{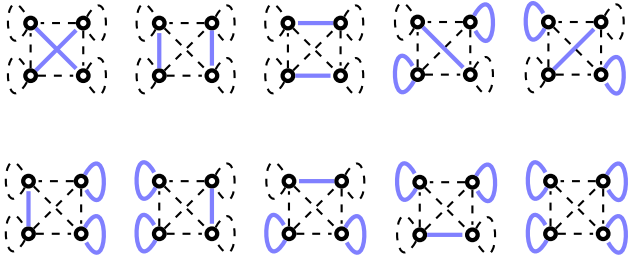}
	\caption{\label{fig:single-pair} Perfect matchings for a complete 4 vertex graph with loops. 
		The first three partitions in the top row correspond to the set PMP$(4)$ and consist of two pairs (edges linking different vertices). The next six SPMs have one pair and two singles (loops).  The last perfect matching consists of four singles (loops).}
\end{figure}
Note that for a given graph size $n$, the set of perfect matchings of a complete graph with loops is significantly larger than the number of perfect matchings of a graph without loops. The ratio of these quantities is exponential in $\sqrt{n}$ (see Ref. \cite{bjorklund2018faster})
\eq{
	\frac{|\text{SPM}(n)|}{|\text{PMP}(n)|}	\sim \frac{\exp\left(\sqrt{n} - \tfrac{1}{4}\right)}{2}, \quad \text{ for } n \gg 1.
} 

An important question concerning a given graph is the number of perfect matchings it has. One possible way to answer this is to iterate over the perfect matchings of a complete graph and at each step check if the given perfect matching of the complete graph is also a matching of the given graph. A simple way to automatize this process  is by constructing the adjacency matrix of the graph. The adjacency matrix $\bm{G}$ of a graph $G=(V,E)$ is a 0-1 matrix that has ${G}_{i,j} = {G}_{j,i}=1$ if, and only if, $(i,j) \in  E$ and 0 otherwise. For the graph in Fig.~\ref{fig:graphs},
(a) the adjacency matrix is
\eq{\label{A1}
\bm{G}^1 = \begin{bmatrix}
	1 & 0 & 0 & 1 & 0 & 0 \\
	0 & 0 & 0 & 1 & 1 & 0 \\
	0 & 0 & 0 & 1 & 1 & 1 \\
	1 & 1 & 1 & 0 & 0 & 0 \\
	0 & 1 & 1 & 0 & 1 & 0 \\
	0 & 0 & 1 & 0 & 0 & 0
\end{bmatrix}.
}
The number of perfect matchings of a graph with/without loops is simply given by the loop hafnian/hafnian of its adjacency matrix
\eq{
\lhaf(\bm{G}) =  \sum_{M \in  \text{SPM}(n)} \prod_{\scriptscriptstyle (i,j) \in  M} {G}_{i,j},\\
\haf(\bm{G}) =  \sum_{M \in
	  \text{PMP}(n)} \prod_{\scriptscriptstyle (i,j) \in  M} {G}_{i,j}.
}
For the graph in Fig (\ref{fig:graphs}) with adjancency matrix in Eq.~(\ref{A1}), we have
\eq{
\haf(\bm{G}^{1})	= 1, \quad \lhaf(\bm{G}^{1})	= 2.
}
The definition of the hafnian and loop hafnian immediately generalizes to weighted graphs, where we assign a real or complex number to the entries of the symmetric matrix $\bm{G}$.
As defined, the loop hafnian and hafnian of an $n \times n$ (complex or real) matrix require $|\text{SPM}(n)| \sim \tfrac{1}{2} e^{\sqrt{n}-1/4} (n-1)!!$ and $|\text{PMP}(n)|= (n-1)!!$ sums respectively. Recently\cite{bjorklund2018faster}, it was shown that this time, both for the loop hafnian and the hafnian, can be cut to $O(n^3 2^{n/2})$, significantly speeding their calculation.

\section{Gaussian bosonic operations}\label{sec:HO}
In this section, we review some basic properties of Gaussian bosonic operations. 
The defining characteristic of these operations is that, when used to transform the set of creation and destruction bosonic operators in the Heisenberg picture, they map them to linear combinations of the same set. Another property of these operations is that, when applied to the $\bm{0}$ boson state, they generate states with Gaussian Wigner functions\cite{weedbrook2012gaussian,serafini2017quantum}. The identities and definitions developed in this section will be used in subsequent sections to map the calculation of FCFs to loop hafnians.

We consider a system of $\ell$ 
bosonic modes with creation and destruction operators $\a_j$, $\ad_j$, satisfying the bosonic commutation relations
\eq{
[\a_j,\ad_k] = \delta_{j,k}, \quad [\a_j,\a_k]=[\ad_j,\ad_k]=0.
}
For each mode, one can also associate Hermitian canonical position $\hat R_j$ and momentum operators $\hat P_j$ defined as
\eq{\label{RandP}
\hat R_{j} &= \frac{1}{\sqrt{2}}\left(\a_{j}+\ad_{j} \right), \quad \hat P_{j} = \frac{1}{\sqrt{2} i}\left(\a_{j}-\ad_{j} \right),
}
that satisfy the usual canonical commutation relations
\eq{
[\hat R_j, \hat P_k ] = 	i \delta_{j,k}, \quad [\hat R_j, \hat R_k ] = [\hat P_j, \hat P_k ] = 0,
}
where, consistent with standard atomic units, we set $\hbar =1$.

Having set up the operator notation let us study certain operations acting on these mode operators. For mode $j$, the displacement operator by amount $\alpha_j \in \mathbb{C}$ is defined as\cite{barnett2002methods}
\seq{
\eq{
	\hat D_j(\alpha_j) &= \hat D_j^\dagger(-\alpha_j)= \exp\left(\alpha_j \ad_j-\alpha^* \a_j \right) \label{no1}	\\
	&=\exp\left(-\tfrac{|\alpha_j|^2}{2} \right) \exp\left(\alpha_j \ad_j \right) \exp\left(-\alpha_j^* \a_j \right).
}
}
When used to transform operators, $\hat D_j(\alpha_j)$  acts as follows:
\seq{
	\label{disp}
	\eq{
	\hat D_j^\dagger(\alpha_j) \a_j \hat D_j(\alpha_j) &= \a_j+\alpha_j,\\
	\hat D_j(\alpha_j) \a_j \hat D_j^\dagger(\alpha_j) &= \a_j-\alpha_j, \label{dispback}\\
	\hat D_j^\dagger(\alpha_j)\hat R_j \hat D_j(\alpha_j) &= \hat R_j+\sqrt{2} \Re(\alpha_j),\\
	\hat D_j^\dagger(\alpha_j) \hat P_j \hat D_j(\alpha_j) &= \hat P_j+\sqrt{2} \Im(\alpha_j).
}
}

The second operation we consider is squeezing. For mode $j$, the squeezing operator by amount $\lambda_j$ is defined as\cite{barnett2002methods} 
\seq{
	\eq{
	\hat S_j(\rr_j)=&\exp\left( \tfrac{\rr_j}{2} \left\{\hat a_j^{\dagger 2}- \a_j^2 \right\}\right) \\
	=&  \exp( \tanh(\rr_j) \hat a_j^{\dagger 2}/2) \label{no2} \\
	& \times \exp\left(-\left(\ad_j \a_j+\tfrac{1}{2} \right) \ln(\cosh(\rr_j))\right) \nonumber\\
	& \times \exp( -\tanh(\rr_j)  \a_j^2/2).   \nonumber
}
}
When used to transform operators, $\hat S$ acts as follows:
\sseq{\label{sq}
	\hat S_j^\dagger(\rr_j) \a_j \hat S_j(\rr_j) &= \a_j \cosh \rr_j +\ad_j \sinh \rr_j ,\\
	\hat S_j^\dagger(\rr_j) \ad_j \hat S_j(\rr_j) &= \ad_j \cosh \rr_j +\a_j \sinh \rr_j,\\
	\hat S_j^\dagger(\rr_j)\hat R_j \hat S_j(\rr_j) &= e^{\rr_j}\hat R_j,\\
	\hat S_j^\dagger(\rr_j) \hat P_j \hat S_j(\rr_j) &= e^{- \rr_j}\hat P_j.
}

The third operation we consider is two-mode squeezing. As the name suggests, this operation acts on two modes, $j$ and $k\neq j$, and it is defined as follows:
\eq{
\hat T_{j,k}(t) = \exp\left(t (\ad_j \ad_k - \a_j \a_k)\right).
}
When acted on the $\bm{0}$ boson state of modes $i,j$, one obtains the two-mode squeezed vacuum state\cite{barnett2002methods}
\eq{
\hat T_{i,j}(t)\ket{0_i,0_j} =\sum_{n=0}^\infty   \frac{\tanh^n (t)}{\cosh (t)}    \ket{n_i,n_j}.	
}
This state has perfect boson-number correlations in the sense that, if one were to measure the probabilities of mode $i$ having $m$ bosons and mode $j$ having $n$, bosons one would find that $p(m,n)=0$ unless $m=n$. Equivalently, if the boson number of mode $j$ is measured and found to be $n$, one immediately knows that mode $k$ has been ``collapsed'' into a state with exactly that same boson number $n$
\eq{\label{TMSVid0}
\left( \id_j \otimes \ket{n_k} \bra{n_k} \right) 	\hat T_{j,k}(t)\ket{0_j,0_k} = \left( \frac{\tanh^n t}{\cosh t}  \right) \ket{n_j,n_k}.
}

Before moving on to describe the last operation, let us introduce some notation related to vectorization. We define multimode versions of the displacement and squeezing operators as follows:
\sseq{
\mathcal{\hat D}(\bm{\alpha}) &= \bigotimes_{j=1}^\ell \hat D_j(\alpha_j),\\
\mathcal{\hat S}(\bm{\lambda}) &= \bigotimes_{j=1}^\ell \hat S_j(\lambda_j),
}
where $\bm{\alpha} \in \mathbb{C}^\ell$ and $\bm{\lambda} \in \mathbb{R}^\ell$. We can also define vectors of operators such as
\sseq{
\R = (\hat R_1, \ldots \hat R_\ell)^T,	\\
\P = (\hat P_1, \ldots \hat P_\ell)^T,	\\
\bm{\a} = (\hat a_1, \ldots \hat a_\ell)^T	.
}
One can easily write vector versions of Eq.~(\ref{disp}) and Eq.~(\ref{sq}), for example,
\seq{\label{heisSD}
	\eq{
\mathcal{\hat D}^\dagger(\bm{\alpha}) \R	\mathcal{\hat D}(\bm{\alpha}) &= \R+\sqrt{2} \Re(\bm{\alpha}), \\
\mathcal{\hat S}^\dagger(\bm{\rr}) \R	\mathcal{\hat S}(\bm{\rr}) &= \exp\left(\bm{\Lambda} \right) \R \label{nudiag},
}
}
where 
\eq{
\bm{\Lambda} = \text{diag}(\bm{\lambda}).
}
We use the notation ``diag'' for the function that, given a vector with entries $\lambda_j$, constructs the matrix $\Lambda_{i,j} = \delta_{i,j} \lambda_j$.

Let us now consider the fourth set of transformations: linear passive transformations $\uu$. These take as parameter a unitary matrix $\bm{U}$ and act as follows on the bosonic operators:
\eq{\label{mapping}
	\uu^\dagger(\bm{U}) \ad_i \uu(\bm{U})=\sum_{l=1}^\ell U^*_{i l}\a_l^\dagger, \quad 
	\uu(\bm{U}) \ad_l \uu^\dagger(\bm{U})=\sum_{i=1}^\ell U_{i l}\a_i^\dagger.
}
If $\bm{U} = \bm{O} \in \mathbb{R}^{\ell \times \ell}$ is real and unitary (i.e. an orthogonal matrix), one can transform the coordinates and momenta as follows:
\eq{\label{heisU}
\uu^\dagger(\bm{O}) \ \R \ \uu(\bm{O})	 = \bm{O} \R,\\
\uu^\dagger(\bm{O}) \ \P \ \uu(\bm{O})	 = \bm{O} \P.
}
An important property of linear passive transformations is that they map the zero boson state to itself
\eq{
\uu(\bm{U}) \ket{\bm{0}} = 	\ket{\bm{0}},
}
where $\ket{\bm{0}}$ is the 0 boson state which is annihilated by all the destruction operators $\a_j \ket{\bm{0}} = 0 \  \forall j$. This is readily seen by noticing that the total particle number operator $\mathcal{\hat N} = \sum_{j=1}^\ell \ad_j \a_j$ is unchanged by transformation via $\uu(\bm{U})$, i.e.
\eq{
\uu^\dagger(\bm{U}) \  \mathcal{\hat N} \  \uu(\bm{U}) = \mathcal{\hat N}.
}
Thus, the operator $\uu(\bm{U})$ does not change the number of particles, hence the name passive.

Having introduced all the Gaussian unitary operations, we now use them to construct more complicated linear transformations of the position operators. We would like to decompose the following Heisenberg picture transformation: 
\eq{\label{genA}
\mathcal{\hat W}^\dagger \Rn \mathcal{\hat W} = \bm{A} \Rn +\bm{d}.	
}
This can be achieved using the recipe from Refs.\cite{huh2015boson,killoran2018continuous}. We first write the singular value decomposition (SVD) of $\bm{A}$ as follows:
\eq{\label{SVD}
\bm{A} = \bm{O}_L \ \bm{L} \ \bm{O}_R^T,
}
where $\bm{O}_{L/R}$ are orthogonal matrices and $\bm{L} = \diag(\bm{l})$ is a diagonal matrix with positive entries.
With the SVD at hand, we simply argue that the following product of Gaussian unitaries will do the job:
\eq{\label{fact}
\mathcal{\hat W}= \mathcal{\hat U}(\bm{O}_R^T) \mathcal{\hat S}(\log(\bm{l})) \mathcal{\hat U}(\bm{O}_L) \mathcal{\hat D}(\bm{d}/\sqrt{2}).
}
This can be readily seen by substituting the last equation into the LHS of Eq.~(\ref{genA}) and then use Eqs.~(\ref{heisSD},\ref{heisU}).

\begin{figure}[!t]
	\includegraphics[width=0.47\textwidth]{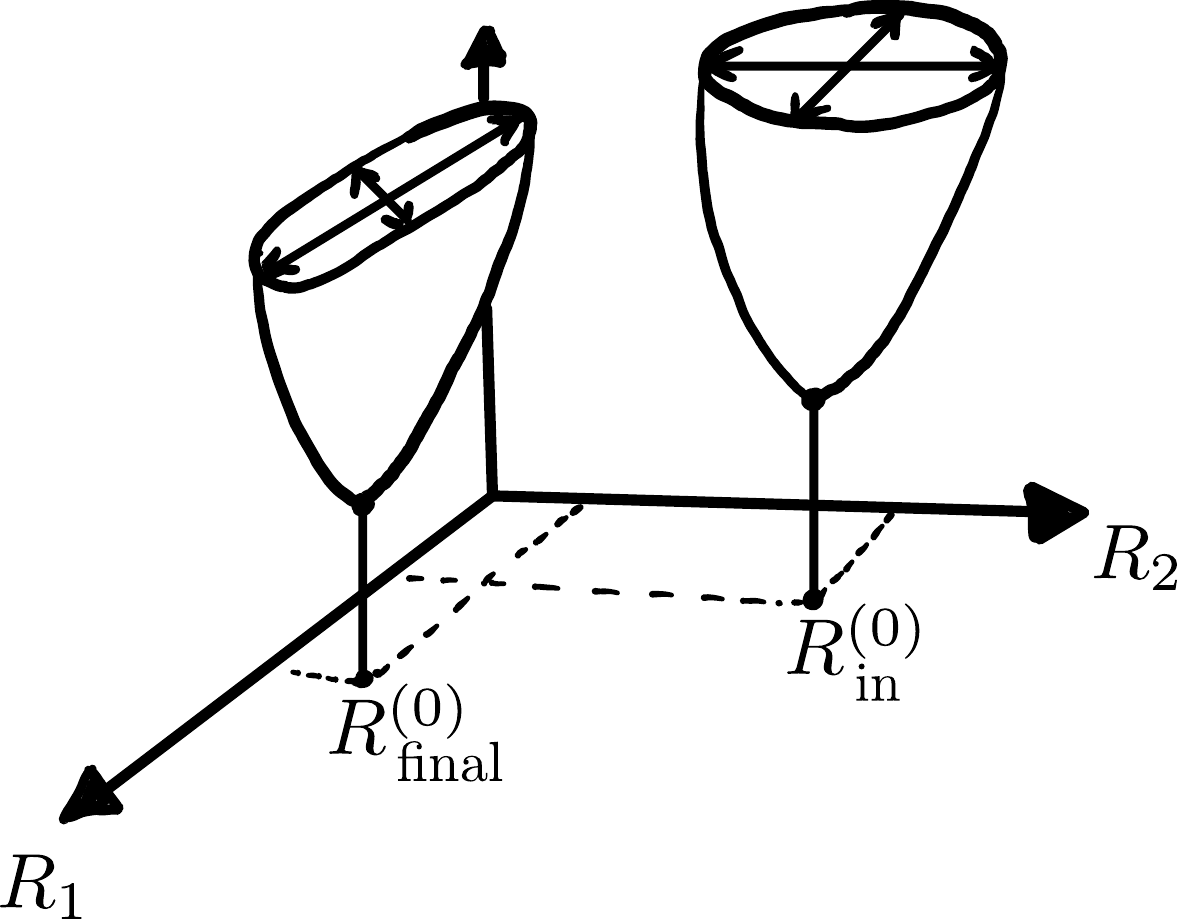}
	\caption{\label{parabolas} Schematic representation of two nuclear energy surfaces; the position of their equilibrium positions $\bm{R}_{\text{\inn/\text{final}}}^{(0)}$, the orientation of their normal axes and their curvatures are different, which leads to a nontrivial Franck-Condon factor. }
\end{figure}
\section{Franck-Condon Factors}\label{sec:FCFs}
In this section we present a review of standard methods to calculate FCFs (see e.g. Huh \cite{huh2011unified} for a thorough discussion) using the notation introduced in the previous section. Furthermore, we show that this quantities are equivalent to the calculation of matrix elements of Gaussian unitaries.

Within first order perturbation theory and in the  Born-Oppenheimer approximation\cite{born1927quantentheorie}, vibronic transitions are dictated by the matrix element 
\eq{\label{matelem1}
\mu_{\out,\inn} = \left\langle \psi_{\out}^e \psi_\out^v  \right| \bm{\hat \mu} \left| \psi_\inn  ^e \psi_\inn ^v \right\rangle,
}
of the dipole moment operator 
\eq{\label{dipole}
\boldsymbol{\hat \mu} = \boldsymbol{\hat \mu}^e + \boldsymbol{\hat \mu}^v,
}
where $\psi_\inn ^e,\psi_\out^e$ and $\psi_\inn ^v, \psi_\out^v$ are the electronic and vibrational wavefunctions of the ground and excited states of the molecule, and in Eq.~(\ref{dipole}) we have separated the dipole operator between its electronic contribution $\boldsymbol{\hat \mu}_e$ and its nuclear contribution $\boldsymbol{\hat \mu}_v$.

Since the electronic dipole operator only acts over the electronic degrees of freedom, we can write Eq.~(\ref{matelem1}) as 
\eq{
\mu_{\out,\inn}&= \braket{\psi^v_\out|\psi^v_\inn} \times  \bm{\mu}^e_{\out,\inn}  
}
where $\bm{\mu}^e_{\out,\inn}$ denotes the electronic transition dipole and, consistent with the FC approximation, we ignore the dependence of the electronic dipole matrix elements on the nuclear coordinates.

The Franck-Condon factor
\eq{
\text{FCF} =\braket{\psi^v_\out|\psi^v_\inn },
}
is the overlap between the vibrational eigenfunctions dwelling in the  excited electronic state potential energy surface and their corresponding eigenfunction for the vibrations in the electronic ground state energy surface.
The vibrational wave functions of the $\ell$ normal modes can be approximated as harmonic oscillator wave functions\cite{atkins2011molecular} obtained by expanding, up to second order, the potential in which the nuclei move around the minima in both the ground and excited state. Thus, for both the ground ($I=\inn$) and excited ($I = \out$) electronic states, we write the nuclear Hamiltonian in atomic units and mass weighted coordinates and momenta $\R, \P$ as
\eq{\label{HI}
\mathcal{\hat H}_{I} =& \tfrac{1}{2}\Pn^T \Pn   +	\tfrac{1}{2} \left(\Rn - \bm{R}_{I}^{(0)} \right)^T \bm{H}_{I} \left(\Rn - \bm{R}_I^{(0)}\right), 
}
where $\bm{H}_I$ is the (real-symmetric) Hessian matrix of the minima associated with the electronic surface $I$.

We can diagonalize the Hessian matrix as
\eq{
\bm{H}_I = \bm{O}_I \bm{\Omega}_I^2	\bm{O}_I^T, 
}
where the matrix $\bm{O}_I$ is an orthogonal matrix giving the directions of the normal modes of vibration and $\bm{\Omega}_I  = \text{diag}(\bm{\omega}_I)$ is a square diagonal matrix containing the normal mode frequencies. 

For each electronic surface, one can diagonalize the Hamiltonian in Eq.~(\ref{HI}) by introducing normal coordinates and momenta according to
\sseq{\label{normalmodes1}
\bm{\hat  R}_I &= \bm{\Omega}^{\tfrac{1}{2}}_I \bm{O}_I^T (\Rn - \bm{R}_I^{(0)}),\\
\bm{\hat  P}_I &= \bm{\Omega}^{-\tfrac{1}{2}}_I \bm{O}_I^T \Pn,
}
in terms of which the Hamiltonian becomes
\eq{\label{Hnormal}
\mathcal{\hat H}_I = \tfrac{1}{2} 	\bm{\hat  P}_I^T \bm{\Omega}_I \bm{\hat  P}_I + \tfrac{1}{2} 	\bm{\hat  R}_I^T \bm{\Omega}_I \bm{\hat  R}_I.
}
Note the appearance of the matrix $\bm{\Omega}_I$ in between the $\bm{\hat P}_I$s. This matrix appears in Eq.~(\ref{Hnormal}) because we have chosen \emph{frequency weighted} (note the factors $\bm{\Omega}_I^{\pm 1/2}$ in Eq.~(\ref{normalmodes1}) ) normal coordinates $\Rn_I$ and momenta $\Pn_I$. This choice of coordinates allows us to write creation and annihilation operators that are not explicitly dependent on the normal frequencies (see Eq.~(\ref{RandP})) 
\eq{
\a_{I,j} = \frac{1}{\sqrt{2}} \left( \Rn_{I,j}+i \Pn_{I,j} \right), \quad \ad_{I,j} = \frac{1}{\sqrt{2}} \left( \Rn_{I,j}-i \Pn_{I,j} \right),
}
in terms of which the Hamiltonian in Eq.~(\ref{Hnormal}) simply becomes
\eq{
\mathcal{H}_I = \sum_{j=1}^\ell (\omega_{I})_{j}\left(\ad_{I,j} \a_{I,j}+\tfrac{1}{2} \right).	
}

The relation connecting the initial and final frequency-weighted normal coordinates can also be written directly:
\sseq{\label{transformation}
\R_\out =& \bm{A} \R_\inn +\bm{d} = \mathcal{\hat W}^\dagger \R_\inn \mathcal{\hat W} ,  \\
\bm{A} =& \left( \bm{\Omega_\out}^{1/2} \bm{O}_\out^T \bm{O}_\inn \bm{\Omega_\inn}^{-1/2} \right) ,\\
\bm{d} =& \left(\bm{\Omega_\out}^{1/2} \bm{O}_\out^T \right) \left(\bm{R}^{(0)}_\inn-\bm{R}^{(0)}_\out \right).
}
The orthogonal matrix $\bm{O}_{D} = \bm{O}_\out^T \bm{O}_\inn $ is the so-called Duschinsky transformation\cite{duschinsky1937importance}, which maps the normal axes of vibration associated to the two electronic surfaces.
The Hilbert space unitary mapping the $\inn$itial and $\out$ operators, $\mathcal{\hat W}$, is the so-called Doktorov transformation\cite{doktorov1975dynamical,doktorov1977dynamical} which can be written exactly in terms of Gaussian unitaries, as in Eq.~(\ref{fact}).

Note that the Doktorov transformation can used to map any operator between the initial and final energy surface, so for example
\eq{\label{atrans}
\a_{\out,j} = 	\mathcal{\hat W}^\dagger  \a_{\inn,j} \mathcal{\hat W},
}
and it can also be used to write the $\bm{0}$ vibrational quanta states of the different energy surfaces in terms of each other
\eq{\label{vacdef}
	\ket{\bm{0}_\out} = \mathcal{\hat W}^\dagger \ket{\bm{0}_\inn}.
}

Having constructed the ground state, we can now construct the excited states by successive application of creation operators 
\eq{\label{fockdef}
\ket{\bm{n}_I} &= 	\bigotimes_{j=1}^{\ell} \frac{\left( \a_{I,j}^\dagger \right)^{n_j}}{\sqrt{n_j!}} \ket{\bm{0}_I},
}
where $\bm{n} = (n_1,\ldots,n_\ell)$ is an $\ell$-dimensional vector of integers that indicates the number of vibrational quanta in each normal mode.
With all these definitions, we can finally go back and write our FCF amplitude as
\seq{
\eq{
\text{FCF}(\bm{n},\bm{m}) &= \braket{\bm{n}_{\out}|\bm{m}_\inn }  \\
&= \bra{\bm{0}_\out}  \bigotimes_{j=1}^{\ell} \frac{\hat a_{\out,j}^{n_j}}{\sqrt{n_j!}}   \ket{\bm{m}_\inn} \label{dec1} \\
&= \bra{\bm{0}_\inn} \mathcal{\hat W} \left( \mathcal{\hat W}^\dagger     \bigotimes_{j=1}^{\ell} \frac{\hat a_{\inn,j}^{n_j}}{\sqrt{n_j!}} \mathcal{\hat W}  \right) \ket{\bm{m}_\inn} \label{dec2}\\
&= \bra{\bm{0}_\inn}  \bigotimes_{j=1}^{\ell} \frac{\hat a_{\inn,j}^{n_j}}{\sqrt{n_j!}} \mathcal{\hat W}   \ket{\bm{m}_\inn} \\
&=\braket{\bm{n}_\inn| \mathcal{W}     |\bm{m}_\inn},
\label{final}
}
}
where in Eq.~(\ref{dec1}) we used Eq.~(\ref{fockdef}) and in Eq.~(\ref{dec2}) we used Eq.~(\ref{vacdef}) and Eq.~(\ref{atrans}).

Since the initial ket and output bra in Eq.~(\ref{final}) refer to the same set of normal coordinates, from now on we drop the ``$\inn$'' label.

With the last equation, we are able to express the FCF as the overlap between two Fock multimode states connected by a (product of) Gaussian unitaries as follows:
\eq{
\text{FCF}(\bm{n}&,\bm{m})= \\
&= \braket{\bm{n}|{\uu(\bm{O}_L^T) \mathcal{\hat S}( \log(\bm{l}))  \uu(\bm{O}_R) \mathcal{\hat D}(\bm{d}/\sqrt{2})} |\bm{m}}\nn\\
&= \braket{\bm{m}|\left(\uu(\bm{O}_L^T) \mathcal{\hat S}( \log(\bm{l}))  \uu(\bm{O}_R) \mathcal{\hat D}(\bm{d}/\sqrt{2})\right)^\dagger |\bm{n}}^*\nn \\
&= \braket{\bm{m}|\mathcal{\hat D}(-\bm{d}/\sqrt{2}) \uu(\bm{O}_R^T)  \mathcal{\hat S}( -\log(\bm{l}))   \uu(\bm{O}_L) |\bm{n}}.\nn
}
In the last equation, we took advantage of the fact that $\braket{\bm{n}|\mathcal{\hat W}|\bm{m}} = \braket{\bm{m}|\mathcal{\hat W}^\dagger|\bm{n}} $, and that the FCFs are real since the arguments of all the Gaussian operators are real matrices or vectors, 
to rewrite the amplitude with the displacement operators appearing ``last''.

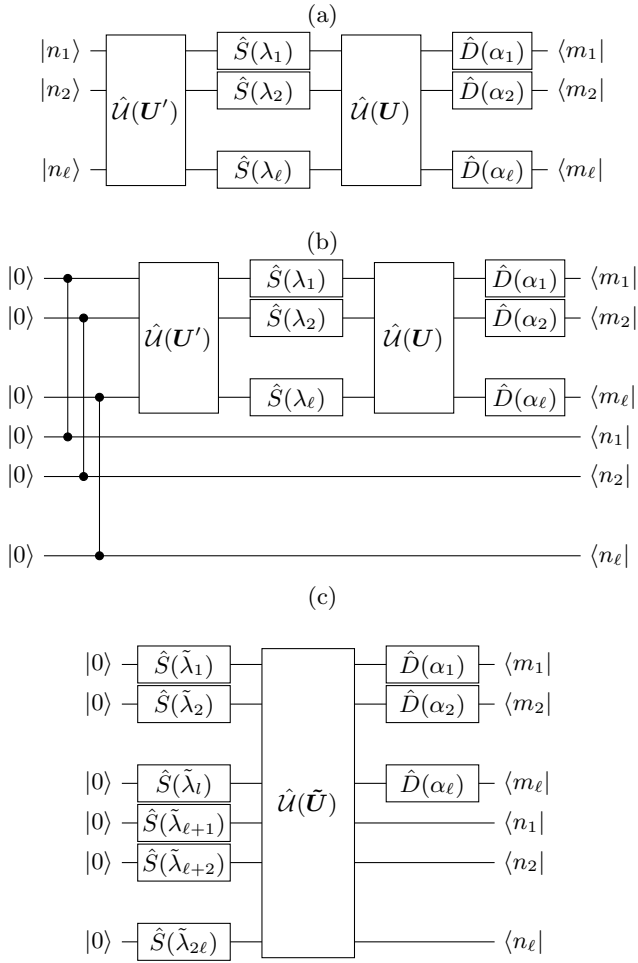
\begin{figure}[!t]
(a) \\	\providecommand{\ket}[1]{\left|#1\right\rangle}
\providecommand{\bra}[1]{\left\langle#1\right|}
\begin{tikzpicture}[scale=1.000000,x=1pt,y=1pt]
\filldraw[color=white] (0.000000, -7.500000) rectangle (173.000000, 52.500000);
\draw[color=black] (0.000000,45.000000) -- (173.000000,45.000000);
\draw[color=black] (0.000000,45.000000) node[left] {$\ket{n_1}$};
\draw[color=black] (0.000000,30.000000) -- (173.000000,30.000000);
\draw[color=black] (0.000000,30.000000) node[left] {$\ket{n_2}$};
\draw[color=white] (0.000000,15.000000) -- (173.000000,15.000000);
\draw[color=white] (0.000000,15.000000) node[left] {$\vdots$};
\draw[color=black] (0.000000,0.000000) -- (173.000000,0.000000);
\draw[color=black] (0.000000,0.000000) node[left] {$\ket{n_\ell}$};
\draw (21.000000,45.000000) -- (21.000000,0.000000);
\begin{scope}
\draw[fill=white] (21.000000, 22.500000) +(-45.000000:21.213203pt and 40.305087pt) -- +(45.000000:21.213203pt and 40.305087pt) -- +(135.000000:21.213203pt and 40.305087pt) -- +(225.000000:21.213203pt and 40.305087pt) -- cycle;
\clip (21.000000, 22.500000) +(-45.000000:21.213203pt and 40.305087pt) -- +(45.000000:21.213203pt and 40.305087pt) -- +(135.000000:21.213203pt and 40.305087pt) -- +(225.000000:21.213203pt and 40.305087pt) -- cycle;
\draw (21.000000, 22.500000) node {$\mathcal{\hat U}(\bm{U}')$};
\end{scope}
\begin{scope}
\draw[fill=white] (65.500000, 45.000000) +(-45.000000:24.748737pt and 9.899495pt) -- +(45.000000:24.748737pt and 9.899495pt) -- +(135.000000:24.748737pt and 9.899495pt) -- +(225.000000:24.748737pt and 9.899495pt) -- cycle;
\clip (65.500000, 45.000000) +(-45.000000:24.748737pt and 9.899495pt) -- +(45.000000:24.748737pt and 9.899495pt) -- +(135.000000:24.748737pt and 9.899495pt) -- +(225.000000:24.748737pt and 9.899495pt) -- cycle;
\draw (65.500000, 45.000000) node {$\hat S(\lambda_1)$};
\end{scope}
\begin{scope}
\draw[fill=white] (65.500000, 30.000000) +(-45.000000:24.748737pt and 9.899495pt) -- +(45.000000:24.748737pt and 9.899495pt) -- +(135.000000:24.748737pt and 9.899495pt) -- +(225.000000:24.748737pt and 9.899495pt) -- cycle;
\clip (65.500000, 30.000000) +(-45.000000:24.748737pt and 9.899495pt) -- +(45.000000:24.748737pt and 9.899495pt) -- +(135.000000:24.748737pt and 9.899495pt) -- +(225.000000:24.748737pt and 9.899495pt) -- cycle;
\draw (65.500000, 30.000000) node {$\hat S(\lambda_2)$};
\end{scope}
\begin{scope}
\draw[fill=white] (65.500000, -0.000000) +(-45.000000:24.748737pt and 9.899495pt) -- +(45.000000:24.748737pt and 9.899495pt) -- +(135.000000:24.748737pt and 9.899495pt) -- +(225.000000:24.748737pt and 9.899495pt) -- cycle;
\clip (65.500000, -0.000000) +(-45.000000:24.748737pt and 9.899495pt) -- +(45.000000:24.748737pt and 9.899495pt) -- +(135.000000:24.748737pt and 9.899495pt) -- +(225.000000:24.748737pt and 9.899495pt) -- cycle;
\draw (65.500000, -0.000000) node {$\hat S(\lambda_\ell)$};
\end{scope}
\draw (110.000000,45.000000) -- (110.000000,0.000000);
\begin{scope}
\draw[fill=white] (110.000000, 22.500000) +(-45.000000:21.213203pt and 40.305087pt) -- +(45.000000:21.213203pt and 40.305087pt) -- +(135.000000:21.213203pt and 40.305087pt) -- +(225.000000:21.213203pt and 40.305087pt) -- cycle;
\clip (110.000000, 22.500000) +(-45.000000:21.213203pt and 40.305087pt) -- +(45.000000:21.213203pt and 40.305087pt) -- +(135.000000:21.213203pt and 40.305087pt) -- +(225.000000:21.213203pt and 40.305087pt) -- cycle;
\draw (110.000000, 22.500000) node {$\mathcal{\hat U}(\bm{U})$};
\end{scope}
\begin{scope}
\draw[fill=white] (152.000000, 45.000000) +(-45.000000:21.213203pt and 9.899495pt) -- +(45.000000:21.213203pt and 9.899495pt) -- +(135.000000:21.213203pt and 9.899495pt) -- +(225.000000:21.213203pt and 9.899495pt) -- cycle;
\clip (152.000000, 45.000000) +(-45.000000:21.213203pt and 9.899495pt) -- +(45.000000:21.213203pt and 9.899495pt) -- +(135.000000:21.213203pt and 9.899495pt) -- +(225.000000:21.213203pt and 9.899495pt) -- cycle;
\draw (152.000000, 45.000000) node {$\hat D(\alpha_1)$};
\end{scope}
\begin{scope}
\draw[fill=white] (152.000000, 30.000000) +(-45.000000:21.213203pt and 9.899495pt) -- +(45.000000:21.213203pt and 9.899495pt) -- +(135.000000:21.213203pt and 9.899495pt) -- +(225.000000:21.213203pt and 9.899495pt) -- cycle;
\clip (152.000000, 30.000000) +(-45.000000:21.213203pt and 9.899495pt) -- +(45.000000:21.213203pt and 9.899495pt) -- +(135.000000:21.213203pt and 9.899495pt) -- +(225.000000:21.213203pt and 9.899495pt) -- cycle;
\draw (152.000000, 30.000000) node {$\hat D(\alpha_2)$};
\end{scope}
\begin{scope}
\draw[fill=white] (152.000000, -0.000000) +(-45.000000:21.213203pt and 9.899495pt) -- +(45.000000:21.213203pt and 9.899495pt) -- +(135.000000:21.213203pt and 9.899495pt) -- +(225.000000:21.213203pt and 9.899495pt) -- cycle;
\clip (152.000000, -0.000000) +(-45.000000:21.213203pt and 9.899495pt) -- +(45.000000:21.213203pt and 9.899495pt) -- +(135.000000:21.213203pt and 9.899495pt) -- +(225.000000:21.213203pt and 9.899495pt) -- cycle;
\draw (152.000000, -0.000000) node {$\hat D(\alpha_\ell)$};
\end{scope}
\draw[color=black] (173.000000,45.000000) node[right] {$\bra{m_1}$};
\draw[color=black] (173.000000,30.000000) node[right] {$\bra{m_2}$};
\draw[color=white] (173.000000,15.000000) node[right] {$\vdots$};
\draw[color=black] (173.000000,0.000000) node[right] {$\bra{m_\ell}$};
\end{tikzpicture} \\
	\bigskip
(b)	\\	\providecommand{\ket}[1]{\left|#1\right\rangle}
\providecommand{\bra}[1]{\left\langle#1\right|}
\begin{tikzpicture}[scale=1.000000,x=1pt,y=1pt]
\filldraw[color=white] (0.000000, -7.500000) rectangle (203.000000, 112.500000);
\draw[color=black] (0.000000,105.000000) -- (203.000000,105.000000);
\draw[color=black] (0.000000,105.000000) node[left] {$\ket{0}$};
\draw[color=black] (0.000000,90.000000) -- (203.000000,90.000000);
\draw[color=black] (0.000000,90.000000) node[left] {$\ket{0}$};
\draw[color=white] (0.000000,75.000000) -- (203.000000,75.000000);
\draw[color=white] (0.000000,75.000000) node[left] {$\vdots$};
\draw[color=black] (0.000000,60.000000) -- (203.000000,60.000000);
\draw[color=black] (0.000000,60.000000) node[left] {$\ket{0}$};
\draw[color=black] (0.000000,45.000000) -- (203.000000,45.000000);
\draw[color=black] (0.000000,45.000000) node[left] {$\ket{0}$};
\draw[color=black] (0.000000,30.000000) -- (203.000000,30.000000);
\draw[color=black] (0.000000,30.000000) node[left] {$\ket{0}$};
\draw[color=white] (0.000000,15.000000) -- (203.000000,15.000000);
\draw[color=white] (0.000000,15.000000) node[left] {$\vdots$};
\draw[color=black] (0.000000,0.000000) -- (203.000000,0.000000);
\draw[color=black] (0.000000,0.000000) node[left] {$\ket{0}$};
\draw (9.000000,105.000000) -- (9.000000,45.000000);
\filldraw (9.000000, 105.000000) circle(1.500000pt);
\filldraw (9.000000, 45.000000) circle(1.500000pt);
\draw (15.000000,90.000000) -- (15.000000,30.000000);
\filldraw (15.000000, 90.000000) circle(1.500000pt);
\filldraw (15.000000, 30.000000) circle(1.500000pt);
\draw (21.000000,60.000000) -- (21.000000,0.000000);
\filldraw (21.000000, 60.000000) circle(1.500000pt);
\filldraw (21.000000, 0.000000) circle(1.500000pt);
\draw (51.000000,105.000000) -- (51.000000,60.000000);
\begin{scope}
\draw[fill=white] (51.000000, 82.500000) +(-45.000000:21.213203pt and 40.305087pt) -- +(45.000000:21.213203pt and 40.305087pt) -- +(135.000000:21.213203pt and 40.305087pt) -- +(225.000000:21.213203pt and 40.305087pt) -- cycle;
\clip (51.000000, 82.500000) +(-45.000000:21.213203pt and 40.305087pt) -- +(45.000000:21.213203pt and 40.305087pt) -- +(135.000000:21.213203pt and 40.305087pt) -- +(225.000000:21.213203pt and 40.305087pt) -- cycle;
\draw (51.000000, 82.500000) node {$\mathcal{\hat U}(\bm{U}')$};
\end{scope}
\begin{scope}
\draw[fill=white] (95.500000, 105.000000) +(-45.000000:24.748737pt and 9.899495pt) -- +(45.000000:24.748737pt and 9.899495pt) -- +(135.000000:24.748737pt and 9.899495pt) -- +(225.000000:24.748737pt and 9.899495pt) -- cycle;
\clip (95.500000, 105.000000) +(-45.000000:24.748737pt and 9.899495pt) -- +(45.000000:24.748737pt and 9.899495pt) -- +(135.000000:24.748737pt and 9.899495pt) -- +(225.000000:24.748737pt and 9.899495pt) -- cycle;
\draw (95.500000, 105.000000) node {$\hat S(\lambda_1)$};
\end{scope}
\begin{scope}
\draw[fill=white] (95.500000, 90.000000) +(-45.000000:24.748737pt and 9.899495pt) -- +(45.000000:24.748737pt and 9.899495pt) -- +(135.000000:24.748737pt and 9.899495pt) -- +(225.000000:24.748737pt and 9.899495pt) -- cycle;
\clip (95.500000, 90.000000) +(-45.000000:24.748737pt and 9.899495pt) -- +(45.000000:24.748737pt and 9.899495pt) -- +(135.000000:24.748737pt and 9.899495pt) -- +(225.000000:24.748737pt and 9.899495pt) -- cycle;
\draw (95.500000, 90.000000) node {$\hat S(\lambda_2)$};
\end{scope}
\begin{scope}
\draw[fill=white] (95.500000, 60.000000) +(-45.000000:24.748737pt and 9.899495pt) -- +(45.000000:24.748737pt and 9.899495pt) -- +(135.000000:24.748737pt and 9.899495pt) -- +(225.000000:24.748737pt and 9.899495pt) -- cycle;
\clip (95.500000, 60.000000) +(-45.000000:24.748737pt and 9.899495pt) -- +(45.000000:24.748737pt and 9.899495pt) -- +(135.000000:24.748737pt and 9.899495pt) -- +(225.000000:24.748737pt and 9.899495pt) -- cycle;
\draw (95.500000, 60.000000) node {$\hat S(\lambda_\ell)$};
\end{scope}
\draw (140.000000,105.000000) -- (140.000000,60.000000);
\begin{scope}
\draw[fill=white] (140.000000, 82.500000) +(-45.000000:21.213203pt and 40.305087pt) -- +(45.000000:21.213203pt and 40.305087pt) -- +(135.000000:21.213203pt and 40.305087pt) -- +(225.000000:21.213203pt and 40.305087pt) -- cycle;
\clip (140.000000, 82.500000) +(-45.000000:21.213203pt and 40.305087pt) -- +(45.000000:21.213203pt and 40.305087pt) -- +(135.000000:21.213203pt and 40.305087pt) -- +(225.000000:21.213203pt and 40.305087pt) -- cycle;
\draw (140.000000, 82.500000) node {$\mathcal{\hat U}(\bm{U})$};
\end{scope}
\begin{scope}
\draw[fill=white] (182.000000, 105.000000) +(-45.000000:21.213203pt and 9.899495pt) -- +(45.000000:21.213203pt and 9.899495pt) -- +(135.000000:21.213203pt and 9.899495pt) -- +(225.000000:21.213203pt and 9.899495pt) -- cycle;
\clip (182.000000, 105.000000) +(-45.000000:21.213203pt and 9.899495pt) -- +(45.000000:21.213203pt and 9.899495pt) -- +(135.000000:21.213203pt and 9.899495pt) -- +(225.000000:21.213203pt and 9.899495pt) -- cycle;
\draw (182.000000, 105.000000) node {$\hat D(\alpha_1)$};
\end{scope}
\begin{scope}
\draw[fill=white] (182.000000, 90.000000) +(-45.000000:21.213203pt and 9.899495pt) -- +(45.000000:21.213203pt and 9.899495pt) -- +(135.000000:21.213203pt and 9.899495pt) -- +(225.000000:21.213203pt and 9.899495pt) -- cycle;
\clip (182.000000, 90.000000) +(-45.000000:21.213203pt and 9.899495pt) -- +(45.000000:21.213203pt and 9.899495pt) -- +(135.000000:21.213203pt and 9.899495pt) -- +(225.000000:21.213203pt and 9.899495pt) -- cycle;
\draw (182.000000, 90.000000) node {$\hat D(\alpha_2)$};
\end{scope}
\begin{scope}
\draw[fill=white] (182.000000, 60.000000) +(-45.000000:21.213203pt and 9.899495pt) -- +(45.000000:21.213203pt and 9.899495pt) -- +(135.000000:21.213203pt and 9.899495pt) -- +(225.000000:21.213203pt and 9.899495pt) -- cycle;
\clip (182.000000, 60.000000) +(-45.000000:21.213203pt and 9.899495pt) -- +(45.000000:21.213203pt and 9.899495pt) -- +(135.000000:21.213203pt and 9.899495pt) -- +(225.000000:21.213203pt and 9.899495pt) -- cycle;
\draw (182.000000, 60.000000) node {$\hat D(\alpha_\ell)$};
\end{scope}
\draw[color=black] (203.000000,105.000000) node[right] {$\bra{m_1}$};
\draw[color=black] (203.000000,90.000000) node[right] {$\bra{m_2}$};
\draw[color=white] (203.000000,75.000000) node[right] {$\vdots$};
\draw[color=black] (203.000000,60.000000) node[right] {$\bra{m_\ell}$};
\draw[color=black] (203.000000,45.000000) node[right] {$\bra{n_1}$};
\draw[color=black] (203.000000,30.000000) node[right] {$\bra{n_2}$};
\draw[color=white] (203.000000,15.000000) node[right] {$\vdots$};
\draw[color=black] (203.000000,0.000000) node[right] {$\bra{n_\ell}$};
\end{tikzpicture}
	\bigskip
(c)	\\	\providecommand{\ket}[1]{\left|#1\right\rangle}
\providecommand{\bra}[1]{\left\langle#1\right|}
\begin{tikzpicture}[scale=1.000000,x=1pt,y=1pt]
\filldraw[color=white] (0.000000, -7.500000) rectangle (141.000000, 112.500000);
\draw[color=black] (0.000000,105.000000) -- (141.000000,105.000000);
\draw[color=black] (0.000000,105.000000) node[left] {$\ket{0}$};
\draw[color=black] (0.000000,90.000000) -- (141.000000,90.000000);
\draw[color=black] (0.000000,90.000000) node[left] {$\ket{0}$};
\draw[color=white] (0.000000,75.000000) -- (141.000000,75.000000);
\draw[color=white] (0.000000,75.000000) node[left] {$\vdots$};
\draw[color=black] (0.000000,60.000000) -- (141.000000,60.000000);
\draw[color=black] (0.000000,60.000000) node[left] {$\ket{0}$};
\draw[color=black] (0.000000,45.000000) -- (141.000000,45.000000);
\draw[color=black] (0.000000,45.000000) node[left] {$\ket{0}$};
\draw[color=black] (0.000000,30.000000) -- (141.000000,30.000000);
\draw[color=black] (0.000000,30.000000) node[left] {$\ket{0}$};
\draw[color=white] (0.000000,15.000000) -- (141.000000,15.000000);
\draw[color=white] (0.000000,15.000000) node[left] {$\vdots$};
\draw[color=black] (0.000000,0.000000) -- (141.000000,0.000000);
\draw[color=black] (0.000000,0.000000) node[left] {$\ket{0}$};
\begin{scope}
\draw[fill=white] (23.500000, 105.000000) +(-45.000000:24.748737pt and 9.899495pt) -- +(45.000000:24.748737pt and 9.899495pt) -- +(135.000000:24.748737pt and 9.899495pt) -- +(225.000000:24.748737pt and 9.899495pt) -- cycle;
\clip (23.500000, 105.000000) +(-45.000000:24.748737pt and 9.899495pt) -- +(45.000000:24.748737pt and 9.899495pt) -- +(135.000000:24.748737pt and 9.899495pt) -- +(225.000000:24.748737pt and 9.899495pt) -- cycle;
\draw (23.500000, 105.000000) node {$\hat S(\tilde{\lambda}_1)$};
\end{scope}
\begin{scope}
\draw[fill=white] (23.500000, 90.000000) +(-45.000000:24.748737pt and 9.899495pt) -- +(45.000000:24.748737pt and 9.899495pt) -- +(135.000000:24.748737pt and 9.899495pt) -- +(225.000000:24.748737pt and 9.899495pt) -- cycle;
\clip (23.500000, 90.000000) +(-45.000000:24.748737pt and 9.899495pt) -- +(45.000000:24.748737pt and 9.899495pt) -- +(135.000000:24.748737pt and 9.899495pt) -- +(225.000000:24.748737pt and 9.899495pt) -- cycle;
\draw (23.500000, 90.000000) node {$\hat S(\tilde{\lambda}_2)$};
\end{scope}
\begin{scope}
\draw[fill=white] (23.500000, 60.000000) +(-45.000000:24.748737pt and 9.899495pt) -- +(45.000000:24.748737pt and 9.899495pt) -- +(135.000000:24.748737pt and 9.899495pt) -- +(225.000000:24.748737pt and 9.899495pt) -- cycle;
\clip (23.500000, 60.000000) +(-45.000000:24.748737pt and 9.899495pt) -- +(45.000000:24.748737pt and 9.899495pt) -- +(135.000000:24.748737pt and 9.899495pt) -- +(225.000000:24.748737pt and 9.899495pt) -- cycle;
\draw (23.500000, 60.000000) node {$\hat S(\tilde{\lambda}_l)$};
\end{scope}
\begin{scope}
\draw[fill=white] (23.500000, 45.000000) +(-45.000000:24.748737pt and 9.899495pt) -- +(45.000000:24.748737pt and 9.899495pt) -- +(135.000000:24.748737pt and 9.899495pt) -- +(225.000000:24.748737pt and 9.899495pt) -- cycle;
\clip (23.500000, 45.000000) +(-45.000000:24.748737pt and 9.899495pt) -- +(45.000000:24.748737pt and 9.899495pt) -- +(135.000000:24.748737pt and 9.899495pt) -- +(225.000000:24.748737pt and 9.899495pt) -- cycle;
\draw (23.500000, 45.000000) node {$\hat S(\tilde{\lambda}_{\ell+1})$};
\end{scope}
\begin{scope}
\draw[fill=white] (23.500000, 30.000000) +(-45.000000:24.748737pt and 9.899495pt) -- +(45.000000:24.748737pt and 9.899495pt) -- +(135.000000:24.748737pt and 9.899495pt) -- +(225.000000:24.748737pt and 9.899495pt) -- cycle;
\clip (23.500000, 30.000000) +(-45.000000:24.748737pt and 9.899495pt) -- +(45.000000:24.748737pt and 9.899495pt) -- +(135.000000:24.748737pt and 9.899495pt) -- +(225.000000:24.748737pt and 9.899495pt) -- cycle;
\draw (23.500000, 30.000000) node {$\hat S(\tilde{\lambda}_{\ell+2})$};
\end{scope}
\begin{scope}
\draw[fill=white] (23.500000, -0.000000) +(-45.000000:24.748737pt and 9.899495pt) -- +(45.000000:24.748737pt and 9.899495pt) -- +(135.000000:24.748737pt and 9.899495pt) -- +(225.000000:24.748737pt and 9.899495pt) -- cycle;
\clip (23.500000, -0.000000) +(-45.000000:24.748737pt and 9.899495pt) -- +(45.000000:24.748737pt and 9.899495pt) -- +(135.000000:24.748737pt and 9.899495pt) -- +(225.000000:24.748737pt and 9.899495pt) -- cycle;
\draw (23.500000, -0.000000) node {$\hat S(\tilde{\lambda}_{2\ell})$};
\end{scope}
\draw (70.500000,105.000000) -- (70.500000,0.000000);
\begin{scope}
\draw[fill=white] (70.500000, 52.500000) +(-45.000000:24.748737pt and 82.731493pt) -- +(45.000000:24.748737pt and 82.731493pt) -- +(135.000000:24.748737pt and 82.731493pt) -- +(225.000000:24.748737pt and 82.731493pt) -- cycle;
\clip (70.500000, 52.500000) +(-45.000000:24.748737pt and 82.731493pt) -- +(45.000000:24.748737pt and 82.731493pt) -- +(135.000000:24.748737pt and 82.731493pt) -- +(225.000000:24.748737pt and 82.731493pt) -- cycle;
\draw (70.500000, 52.500000) node {$\mathcal{\hat U}(\bm{\tilde U})$};
\end{scope}
\begin{scope}
\draw[fill=white] (117.500000, 105.000000) +(-45.000000:24.748737pt and 9.899495pt) -- +(45.000000:24.748737pt and 9.899495pt) -- +(135.000000:24.748737pt and 9.899495pt) -- +(225.000000:24.748737pt and 9.899495pt) -- cycle;
\clip (117.500000, 105.000000) +(-45.000000:24.748737pt and 9.899495pt) -- +(45.000000:24.748737pt and 9.899495pt) -- +(135.000000:24.748737pt and 9.899495pt) -- +(225.000000:24.748737pt and 9.899495pt) -- cycle;
\draw (117.500000, 105.000000) node {$\hat D(\alpha_1)$};
\end{scope}
\begin{scope}
\draw[fill=white] (117.500000, 90.000000) +(-45.000000:24.748737pt and 9.899495pt) -- +(45.000000:24.748737pt and 9.899495pt) -- +(135.000000:24.748737pt and 9.899495pt) -- +(225.000000:24.748737pt and 9.899495pt) -- cycle;
\clip (117.500000, 90.000000) +(-45.000000:24.748737pt and 9.899495pt) -- +(45.000000:24.748737pt and 9.899495pt) -- +(135.000000:24.748737pt and 9.899495pt) -- +(225.000000:24.748737pt and 9.899495pt) -- cycle;
\draw (117.500000, 90.000000) node {$\hat D(\alpha_2)$};
\end{scope}
\begin{scope}
\draw[fill=white] (117.500000, 60.000000) +(-45.000000:24.748737pt and 9.899495pt) -- +(45.000000:24.748737pt and 9.899495pt) -- +(135.000000:24.748737pt and 9.899495pt) -- +(225.000000:24.748737pt and 9.899495pt) -- cycle;
\clip (117.500000, 60.000000) +(-45.000000:24.748737pt and 9.899495pt) -- +(45.000000:24.748737pt and 9.899495pt) -- +(135.000000:24.748737pt and 9.899495pt) -- +(225.000000:24.748737pt and 9.899495pt) -- cycle;
\draw (117.500000, 60.000000) node {$\hat D(\alpha_\ell)$};
\end{scope}
\draw[color=black] (141.000000,105.000000) node[right] {$\bra{m_1}$};
\draw[color=black] (141.000000,90.000000) node[right] {$\bra{m_2}$};
\draw[color=white] (141.000000,75.000000) node[right] {$\vdots$};
\draw[color=black] (141.000000,60.000000) node[right] {$\bra{m_\ell}$};
\draw[color=black] (141.000000,45.000000) node[right] {$\bra{n_1}$};
\draw[color=black] (141.000000,30.000000) node[right] {$\bra{n_2}$};
\draw[color=white] (141.000000,15.000000) node[right] {$\vdots$};
\draw[color=black] (141.000000,0.000000) node[right] {$\bra{n_\ell}$};
\end{tikzpicture}

		\caption{\label{fig:circ} In (a) we show the circuit corresponding to the amplitude $\nu$ in Eq.~(\ref{mudef}). In (b) we show the equivalent circuit using two-mode squeezed vacuum states to prepare the initial input state as in Eq.~(\ref{murewritten}). The vertical lines with dots at the ends are used to indicate two-mode squeezing operators $\hat T_{i,i+\ell}$ between modes $i$ and $i+\ell$.  In (c) we show an equivalent circuit to the one in (b) after using the Bloch-Messiah decomposition.}
\end{figure}

The results presented, originally derived by Huh \ea \cite{huh2015boson}, show very generally that a FCF is a special case of the following matrix elements:
\eq{\label{mudef}
	\nu = \braket{\bm{m}|\mathcal{\hat D}(\bm{\alpha}) \uu(\bm{U}) \s (\bm{\lambda}) \uu(\bm{U}') |\bm{n}}.
}
This amplitude, represented as a circuit in Fig.~\ref{fig:circ} (a), has a simple interpretation in quantum optics: the ket $\ket{\bm{n}}$ is a (multimode) Fock state that is sent into an interferometer that implements the linear passive transformation $\uu(\bm{U}')$. Then, each output mode $j$  of this interferometer is squeezed by an amount $\lambda_{j}$ and sent to another interferometer that implements the linear passive transformation $\uu(\bm{U})$. Finally, each mode receives a displacement by amount $\alpha_j$.
The quantity $|\nu|^2$ is simply the probability that, when the state just described is measured with PNRD, it is found to have $m_j$ bosons in mode $j$ for all the $\ell$ modes. Thus we say that the state $\mathcal{\hat D}(\bm{\alpha}) \uu(\bm{U}) \s (\bm{\lambda}) \uu(\bm{U}') \ket{\bm{n}}$ is ``measured'' in the bra $\bra{\bm{m}}$ containing $m_j$ photons in mode $j$.

As previously done by Huh and Yung \cite{huh2017vibronic} for FCFs, let us rewrite this matrix element as the matrix element of a system with twice as many modes in which the initial state has zero bosons and one applies an enlarged Gaussian linear transformation. We double the number of modes from $\ell$ to two $2\ell$ and write boson number kets (bras) for these systems as $\ket{\bm{m},\bm{n}}$ $(\bra{\bm{m},\bm{n}})$ where $\bm{m}$ is a vector of $\ell$ integers identifying the number of quanta of the first $\ell$ modes (which are the modes of interest), and similarly, $\bm{n}$ is another vector with $\ell$ integers specifying the number of quanta of the ancillary modes. 
The original matrix elements of interest can always be written as
\eq{\label{munn}
\nu =& \braket{\bm{m,n}| \left( \mathcal{\hat D}(\bm{\alpha})  \otimes \id \right) \left( \uu(\bm{U}) \otimes \id \right) \nn \\
	&\quad \quad \times \left( \s (\bm{\lambda}) \otimes \id \right) \left( \uu(\bm{U}') \otimes \id \right) |\bm{n,n}}	\\
 =& \braket{\bm{m}|\mathcal{\hat D}(\bm{\alpha}) \uu(\bm{U}) \s (\bm{\lambda}) \uu(\bm{U}') |\bm{n}} \times \braket{\bm{n}|\mathbb{\hat I} |\bm{n}}.
}
In the last equation, we tensored the unitaries of the first $\ell$ modes with identity operators in the other $\ell$ ancillary modes.

Now, note that we can write the following identity:
\eq{\label{TMSVid}
\left(\id \otimes   \ket{\bm{n}} \bra{\bm{n}} \right)	\mathcal{\hat T}(\bm{t}) \ket{\bm{0},\bm{0}} = \left( \prod_{k=1}^\ell \frac{\tanh^{n_k} t_j}{\cosh t_j} \right) \ket{\bm{n},\bm{n}},
}
where $\bm{t} = (t_1,\ldots,t_\ell)$ is a vector of (two-mode) squeezing parameters and 
\eq{
\mathcal{\hat T}(\bm{t})	 = \bigotimes_{k=1}^\ell \hat T_{k,\ell+k}(t_k),
}
is a product of two-mode squeezing operations between mode $j$ from the original set of modes and mode $j+\ell$ from the ancillary modes. 
The identity in Eq.~(\ref{TMSVid}) just says that the two-mode squeezed vacuum is a state with perfect number correlations between the two modes being squeezed, and is just a multimode version of Eq.~(\ref{TMSVid0}). Using this identity, we can rewrite $\nu$ in Eq.~(\ref{munn}) as
\eq{\label{murewritten}
\nu = &\frac{1}{\left(\prod_{j=1}^\ell \frac{\tanh^{n_j} t_j}{\cosh t_j} \right)} \bra{\bm{m,n}} \left( \mathcal{\hat D}(\bm{\alpha})  \otimes \id \right) \left( \uu(\bm{U}) \otimes \id \right)  \nn \\
& \quad \quad  \quad  \times \left( \s (\bm{\lambda}) \otimes \id \right) \left( \uu(\bm{U}') \otimes \id \right) \mathcal{\hat T}(\bm{t}) \ket{\bm{0,0}}.	
}
This is represented schematically in Fig.~\ref{fig:circ} (b).
We note that the operation 
\eq{
\mathcal{Q} =  \left( \uu(\bm{U}) \otimes \id \right) \left( \s (\bm{\lambda}) \otimes \id \right)  \left( \uu(\bm{U}') \otimes \id \right) \mathcal{\hat T}(\bm{t})
}
is a Gaussian operation and it admits a Bloch-Messiah decomposition\cite{bloch1975canonical,serafini2017quantum} as follows:
\eq{
\mathcal{Q} = \uu(\bm{\tilde U})	\s(\bm{\tilde \lambda}) \uu(\bm{\tilde U}'),
}
where we use the tildes in the arguments of the $\uu$s and $\s$ to indicate that they act in $2\ell$ modes, and thus $\bm{\tilde U}$ and $\bm{\tilde U}'$ are $2\ell \times 2\ell$ unitary matrices and $\bm{\tilde \lambda}$ is a $2\ell$ real vector. With this new notation, we finally write
\eq{
\nu &= 	\braket{\bm{m}|\mathcal{\hat D}(\bm{\alpha}) \uu(\bm{U}) \s (\bm{\lambda}) \uu(\bm{U}') |\bm{n}} \nn \\
&= R \times \bra{\bm{m,n}} \mathcal{\hat D}(\bm{\tilde \alpha})  \ \uu(\bm{\tilde U})	\s(\bm{\tilde \lambda}) \uu(\bm{\tilde U}') \ket{\bm{0,0}} \nn \\
&= R \times \bra{\bm{m,n}} \mathcal{\hat D}(\bm{\tilde \alpha})  \ \uu(\bm{\tilde U})	\s(\bm{\tilde \lambda}) \ket{\bm{0,0}} \\
&=R \times \bra{\bm{p}} \mathcal{\hat D}(\bm{\tilde \alpha})  \ \uu(\bm{\tilde U})	\s(\bm{\tilde \lambda}) \ket{\bm{0}} 
}
where
\eq{
\bm{p} & = (m_1,m_2,\ldots, m_\ell,n_1,n_2,\ldots,n_\ell),	\\
\bm{\tilde \alpha} &= (\alpha_1,\alpha_2,\ldots,\alpha_\ell,\underbrace{0,0,\ldots,0}_{\ell \text{ times }}), \\
R &= \frac{1}{\prod_{j=1}^\ell \frac{\tanh^{n_j} t_j}{\cosh t_j}}	.
}

By using the Bloch-Messiah decomposition, we have turned the calculation of a matrix element between $\bra{\bm{m}}$ and $\ket{\bm{n}}$ into one of the form $\bra{\bm{m},\bm{n}} \equiv \bra{\bm{p}}$ and $\ket{\bm{0},\bm{0}}\equiv \ket{\bm{0}}$. The result of this mapping is the circuit illustrated in Fig.~\ref{fig:circ} (c). 

Without loss of generality, we will focus only on calculations of matrix elements of the latter form, where the initial ket is always a state with zero bosons in a possibly enlarged set of modes; any nonzero element of $\bm{n}$ in the initial bra can be moved to the final ket by adding one extra mode to the system, using an appropriate two-mode squeezing operation and renormalizing by the prefactor $R$. 

Note that in principle, one can pick any nonzero $t_j$ for the two-mode squeezing operation. In practice, however, one would like to maximize $R$; this is achieved by setting 
\eq{
\sinh^2 t_j = n_j.
}

\section{Gaussian state amplitudes in the Fock basis} \label{sec:amplitudes}
In the last section we reduced the FCF to the calculation of the matrix elements in the Fock basis of a Gaussian state obtained by applying squeezing, passive linear transformations and displacements to the $\bm{0}$ boson state. We will then investigate the following amplitude
\eq{\label{amp1}
	\nu = R \times  \braket{\bm{p}|\mathcal{\hat D}(\ta) \uu(\tu) \s (\bm{\tilde \lambda}) |\bm{0}}.
}
We will show that this amplitude is proportional to a loop hafnian of a certain matrix that is a function of $\bm{\tilde U}, \bm{\tilde \lambda}$ and $\bm{\tilde \alpha}$.
The strategy that will be followed consist in using operator ordering identities such as equations (\ref{no1},\ref{no2}) to write the amplitude $\nu$ as follows:
\eq{\label{amplitude}
	\nu &= R \ T \times \braket{\bm{0}|   \left( \bigotimes_{j=1}^{2\ell} \a_j^{p_j} \right) \exp(\hat A) \exp(\hat C)|\bm{0}},
}	
where we define 
\seq{
	\eq{	
	\hat A&=\frac{1}{2} \sum_{k,m}^{2 \ell} B_{kj} \hat a_m^{\dagger } \hat a_k^{\dagger },\\
	\hat C&= \sum_{j=1}^{2\ell} \zeta_j \ad_j ,\\
		T&= \frac{\exp\left(-\tfrac{1}{2} \left\{ |\bm{\tilde{\alpha}}|^2 - \bm{\tilde{\alpha}}^\dagger \bm{B} \bm{\tilde{\alpha}}^*\right\}  \right)}{\sqrt{\prod_{j=1}^{2 \ell} \left(p_j! \cosh(\tilde{\lambda}_j) \right)}} \\
	\bm{B}&= \tu \tanh(\tl) \tu^T , \\
	\bm{\zeta}&=\ta - \bm{B}\ta^*.
}
}
Note that the operators $\hat A$ and $\hat C$ are second and first degree polynomials of the creation  operators $\ad_j$ only.
The steps leading from Eq. (\ref{amp1}) to Eq. (\ref{amplitude}) are mathematically straightforward but quite lengthy and thus we present them in detail in Appendix \ref{app:reordering}. We remark that these operator reordering results are also useful in the calculation of matrix elements of pure Gaussian states in the coherent state basis (see Appendix \ref{app:coh}).

As a second step, detailed in Sec. \ref{rulem}, we will  map the modes in which more than a single boson is measured ($p_j>1$) to $p_j$ ancillary modes where only a single boson is measured.
In the third part, detailed in Sec. \ref{contract}, we will use the fact that the amplitude we are looking for has two states of definite number of bosons in either side of the two exponentials in Eq.~(\ref{amplitude}) to pick specific terms that connect the initial state with $\bm{0}$ bosons to a state with 
\eq{
	P = \sum_{j=1}^{2\ell} p_j	 =  \underbrace{\sum_{j=1}^{\ell} m_j}_{=M}	+\underbrace{\sum_{j=1}^{\ell} n_j}_{=N},
}
bosons.
Having determined these terms, we expand them using the multinomial theorem and normal order the resulting expression to show that the probability amplitude we are looking for is indeed a loop hafnian.
We summarize the whole procedure in Sec. \ref{summary}.

In Appendix \ref{app:hamilton} we show how our results, which give directly probability amplitudes and do not require the use of phase space methods, reduce in the appropriate limit to the results by Hamilton \ea \cite{hamilton2017gaussian}

\subsection{Multiple bosons in the same mode}\label{rulem}
We would like to write the amplitude in Eq. (\ref{amplitude}) in a slightly different way, in which to the right of $\bra{\bm{0}}$, there are only destruction operators raised to the power of 1.
We will then rewrite the product $\pp$ in terms of a new set of operators while redefining $\hat A$ and $\hat C$.

We define first the multiset $S_{\bm{p}}$ obtained as follows
\eq{
	S_{\bm{p}} = \{\underbrace{1,\ldots,1}_{p_1 \text{ times}}, \underbrace{2,\ldots,2}_{p_2 \text{ times}} \ldots, \underbrace{2\ell,\ldots,2\ell}_{p_{2\ell} \text{ times}} \}	.
}
For instance, if $\bm{p}=(1,3,0,2)$, then
\eq{
	S_{(1,3,0,2)} = \{1,2,2,2,4,4\}	.
}
Note that the number of elements in the multiset $|S_{\bm{p}}| =\sum_{k=1}^\ell p_k \equiv P$.
Now let us construct $\bm{B}^{(\bm{p})}\in\mathbb{C}^{P\times P}$ as the matrix formed by indexing elements of $\bm{B}$ according to the multiset $S_{\bm{p}}$. More precisely, the $j^\text{th}$ row and column of $\bm{B}$ are repeated $p_j$ times (or dropped when $p_j=0$) when forming $\bm{B}^{(\bm{p})}$.
For example if
\eq{
	\bm{B} = 	\begin{bmatrix}
		B_{1,1} & B_{1,2} & B_{1,3} & B_{1,4}\\
		B_{2,1} & B_{2,2} & B_{2,3} & B_{2,4}\\
		B_{3,1} & B_{3,2} & B_{3,3} & B_{3,4}\\
		B_{4,1} & B_{4,2} & B_{4,3} & B_{4,4}
	\end{bmatrix},
}
then
\eq{
	\bm{B}^{(1,3,0,2)} = \left[\begin{array}{c|ccc||cc}
		B_{1,1} & B_{1,2} & B_{1,2} & B_{1,2} & B_{1,4} & B_{1,4}\\
		\hline
		B_{2,1} & B_{2,2} & B_{2,2} & B_{2,2} & B_{2,4} & B_{2,4}\\
		B_{2,1} & B_{2,2} & B_{2,2} & B_{2,2} & B_{2,4} & B_{2,4}\\
		B_{2,1} & B_{2,2} & B_{2,2} & B_{2,2} & B_{2,4} & B_{2,4}\\
		\hline
		\hline
		B_{4,1} & B_{4,2} & B_{4,2} & B_{4,2} & B_{4,4} & B_{4,4}\\
		B_{4,1} & B_{4,2} & B_{4,2} & B_{4,2} & B_{4,4} & B_{4,4}
	\end{array}	\right].
}
The first row and column of $\bm{B}$ is repeated only once in $\bm{B}^{(1,3,0,2)}$, the second one is repeated three times, $p_2=3$, the third one does not appear since  $p_3=0$, and the last one is repeated twice, $p_4=2$. Also, note that $\bm{B}^{(\bm{p})}$ is symmetric if $\bm{B}$ is symmetric.

In a similar manner, we define $\bm{\zeta}^{(\bm{p})} \in \mathbb{C}^P$ as the vector formed by indexing the elements of $\bm{\zeta}$ according to the multiset $S_{\bm{p}}$. For example:
\eq{
	\bm{\zeta} &= [\zeta_1,\zeta_2,\zeta_3,\zeta_4],	\\
	\bm{\zeta}^{(1,3,0,2)} &= [\zeta_1,\zeta_2,\zeta_2,\zeta_2,\zeta_4,\zeta_4].
}
We now argue that
\eq{\label{mapping}
	&\braket{\bm{0}|\pp \exp(\hat A) \exp(\hat C) |\bm{0}	} \\
	&\quad = \bra{\bm{0}} 	\left(\bigotimes_{j=1}^M \b_{j} \right)  \exp(\hat D) \exp(\hat E) \ket{\bm{0}},
}
where
\seq{\label{DandE}
	\eq{
		\hat D &= \frac{1}{2} \sum_{j,k=1}^P \left(\bm{B}^{(\bm{p})}\right)_{jk} \bd_j \bd_k,\\
		\hat E &= \sum_{j=1}^P \left(\bm{\zeta}^{(\bm{p})}\right)_j \bd_j,\\
		[\b_i,\bd_j] &= \delta_{i,j}, \quad [\b_i,\b_j]=[\bd_i,\bd_j] = 0.
	}
}
To justify this, first let us note that, in the case where $p_j = 1 \ \forall j$, we have done nothing but rename the destruction operators $\a \to \b$. In the case where  $p_j \leq 1 \ \forall j$ the only thing we have to do is to remove the operators $\a_j^\dagger$ that will be projected into $0$ bosons. This is justified by noticing that, if mode $j$ has $p_j=0$, then
\eq{
	&\bra{p_1,\ldots, 0_j,\ldots p_{2\ell} } \exp\left(\sum_{k=1}^{2\ell} A_{k,j}\ad_k \ad_j + \frac{1}{2} \a^{\dagger 2}_j +\zeta_j \ad_j \right) 	\nn \\
	&=\bra{p_1,\ldots, 0_j,\ldots p_{2\ell} },
}
since the only term in the Taylor expansion of $\exp(\ldots)$ in which no bosons are created is the first order one, which is equal to the identity $\id$; one might as well simply ignore (drop) the rows and columns of $\bm{B}$ and the elements of $\bm{\zeta}$ where the mode $j$ is found to have zero bosons.

Finally, for the case of $p_j>1$, the mapping in Eq.~(\ref{mapping}) is essentially splitting multiple bosons being measured in a single mode $\bra{\bm{0}} \a_j^{p_j}$ into $p_j$ single bosons being measured into $p_j$ different modes that have the same joint amplitude of pair creation $B_{i,j}$ with respect to the other modes. This is precisely what is achieved by constructing the matrix $\bm{B}^{(\bm{p})}$. Thus going to our four mode example with $\bm{p} = (1,0,3,2)$, we have that the following amplitude
\eq{
	\bra{\bm{0}} \left(\a_1 \a_2^3 \a_4^2 \right) \exp(\hat A) \exp(\hat B) \ket{\bm{0}	} ,
}
with 
\eq{
	\hat A &=\begin{bmatrix}
		\ad_1 , \ad_2, \ad_3, \ad_4
	\end{bmatrix} \begin{bmatrix}
	B_{1,1} & B_{1,2} & B_{1,3} & B_{1,4}\\
	B_{2,1} & B_{2,2} & B_{2,3} & B_{2,4}\\
	B_{3,1} & B_{3,2} & B_{3,3} & B_{3,4}\\
	B_{4,1} & B_{4,2} & B_{4,3} & B_{4,4}
\end{bmatrix}
\begin{bmatrix}
	\ad_1\\ \ad_2\\ \ad_3 \\  \ad_4
\end{bmatrix}, \\
\hat C &= \begin{bmatrix}
	\zeta_1 , \zeta_2 , \zeta_3 , \zeta_4
\end{bmatrix}
\begin{bmatrix}
	\ad_1 \\ \ad_2 \\ \ad_3 \\ \ad_4
\end{bmatrix},
}
equals
\eq{
	\bra{\bm{0}} \left(\b_1 \b_2 \b_3 \b_4 \b_5 \b_6 \right) \exp(\hat D) \exp(\hat E) \ket{\bm{0}	} ,	
}
where
\eq{
	\hat D &= 
	\begin{bmatrix}
		\bd_1, \bd_2, \bd_3,\bd_4,\bd_5,\bd_6
	\end{bmatrix} \\
	&\times 
	\left[\begin{array}{c|ccc||cc}
		B_{1,1} & B_{1,2} & B_{1,2} & B_{1,2} & B_{1,4} & B_{1,4}\\
		\hline
		B_{2,1} & B_{2,2} & B_{2,2} & B_{2,2} & B_{2,4} & B_{2,4}\\
		B_{2,1} & B_{2,2} & B_{2,2} & B_{2,2} & B_{2,4} & B_{2,4}\\
		B_{2,1} & B_{2,2} & B_{2,2} & B_{2,2} & B_{2,4} & B_{2,4}\\
		\hline
		\hline
		B_{4,1} & B_{4,2} & B_{4,2} & B_{4,2} & B_{4,4} & B_{4,4}\\
		B_{4,1} & B_{4,2} & B_{4,2} & B_{4,2} & B_{4,4} & B_{4,4}
	\end{array}	\right]
	\begin{bmatrix}
		\bd_1 \\ \bd_2 \\ \bd_3 \\ \bd_4 \\ \bd_5 \\ \b_6
	\end{bmatrix},\\
	\hat E &= 
	\begin{bmatrix}
		\zeta_1 , \zeta_2 , \zeta_2 , \zeta_2 , \zeta_4 , \zeta_4
	\end{bmatrix}
	\begin{bmatrix}
		\bd_1 \\ \bd_2 \\ \bd_3 \\ \bd_4 \\ \bd_5 \\ \b_6
	\end{bmatrix}.
}
This rule effectively replaces modes where $p_j>1$ bosons are destroyed to $p_j$ independent modes where only a single boson is destroyed. For our 4 mode example
\sseq{
	\a_1 &\underset{p_1=1}{\to} \b_1, \\
	\a_2^3 &\underset{p_2=3}{\to} \b_2,\b_3,\b_4, \\
	\a_3 &\underset{p_3=0}{\to} \{\},	\\
	\a_4 &\underset{p_4=2}{\to} \b_5,\b_6, 
}
where we use $\{\}$ to indicate that $\a_3$ is discarded.

\subsection{Contracting the expression by normal ordering}\label{contract}
We have turned the calculation of the amplitude $\nu$ into the following
\eq{
	\nu = R \ T \times \braket{\bm{0}| \left(\bigotimes_{l=1}^P \b_l \right) \exp\left(\hat D \right) \exp\left(\hat E \right)| \bm{0}}	
}
where $\hat D$  and $\hat E$ are defined in Eq.~(\ref{DandE}). Note furthermore that one can write
\eq{
	\hat D \equiv& \frac{1}{2} \sum_{jk=1}^P \left(\bm{B}^{(\bm{p})}\right)_{j,k}  \bd_j \bd_k \\
	=& \frac{1}{2} \sum_{j=1}^P \left(\bm{B}^{(\bm{p})}\right)_{jj} \bd_j \bd_j + \sum_{j>k}^P \left(\bm{B}^{(\bm{p})}\right)_{j,k} \bd_j \bd_k
}
and, finally, since we are only considering a single boson being destroyed in the modes $\b$, one can neglect the term $\frac{1}{2} \sum_{j=1}^P \left(\bm{B}^{(\bm{p})}\right)_{j,j} \bd_j \bd_j$ since there will never be two $\b_j$s to destroy the pairs of bosons in the same mode this term creates.

The bra $\bra{\bm{0}} \left(\bigotimes_{l=1}^P \b_l \right)$ contains exactly $P$ single bosons, hence, trivially
\eq{
	\bra{\bm{0}}\ppp \hat \Pi_P =  \bra{\bm{0}}\ppp,
} 
where $\hat \Pi_{P}$ is the projector onto the subspace with $P$ bosons.
We can use the last equation to write
\eq{
	\nu &= R \ T  \times \braket{\bm{0}|\ppp \exp(\hat D) \exp(\hat E) |\bm{0}	}	\\
	&=R \ T \times \braket{\bm{0}|\ppp  \hat \Pi_P  \exp(\hat D) \exp(\hat E) |\bm{0}	}.
}
Now we simplify $\hat \Pi_P \exp(\hat D) \exp(\hat E) \ket{\bm{0}	}$ by expanding the exponentials in the RHS of Eq.~(\ref{amplitude})
\eq{
	\exp(\hat D) \exp(\hat E) = \left( \sum_{q=0}^\infty \frac{\hat D^q}{q!} \right) \left(	\sum_{q'=0}^\infty \frac{\hat E^{q'}}{q'!} \right),
}
where each power of the operators $\hat E^q/q!$ or $\hat{E}^{q'}/q'!$ can be expanded using the multinomial theorem:
\eq{\label{multinomial}
	&\frac{(x_1 + x_2  + \cdots + x_m)^n}{n!} \\
	&\quad \quad = \quad \sum_{k_1+k_2+\cdots+k_m=n} \frac{1}{k_1!\, k_2! \cdots k_m!}
	\prod_{t=1}^m x_t^{k_t} . \nonumber
}
This identity, which is valid for complex numbers, also holds for a set of operators that commute, like all the $\bd_j$. Using this observation, it becomes clear that one needs to consider only all the $q$ and $q'$, such that
\eq{\label{totph}
P = 2q+q',
}
where $q$ is the number of pairs of bosons being created by $\hat D$ and $q'$ is the number of single bosons being created by $\hat E^{q'}$; the factor of two in front of $q$ appears precisely because the operator $\hat{D}$ creates two bosons at a time, whereas $\hat E$ creates only singles.
With this observation, we find
\eq{
	&\hat \Pi_P \exp(\hat D) \exp(\hat E) \ket{\bm{0}	} \\
	&=
	\begin{cases}
		\sum_{q=0}^{L} \frac{\hat D^{(L-q)}}{(L-q)!} \frac{\hat E^{2q}}{(2q)!} \ket{\bm{0}}, & L = \frac{P}{2} , \ P \text{ even,}\\
		& \\
		\sum_{q=0}^{L} \frac{\hat D^{(L-q)}}{(L-q)!} \frac{\hat E^{2q+1}}{(2q+1)!} \ket{\bm{0}}, & L=\frac{P-1}{2}, \ P   \text{ odd.} \\
	\end{cases} 
	\nn
}

\subsubsection{Contractions: An example with four modes}
Before embarking on a full calculation, let us look at the case of four modes to gain some intuition. Referring to Eq.~(\ref{totph}), we have $P=4$ and we need to consider the following cases 
\eq{
	2q=4, \ q'=0;\\
	2q=2, \ q'=2; \\
	2q=0, \ q'=4.
}
Let's start with the first case. We only need to consider $2q=4$, for which we have
\eq{\label{example4}
	\frac{\hat D^2}{2!} \frac{\hat E^0}{0!} = 	&\frac{1}{2}\big(\bp_{1,2}\bd_1 \bd_2+\bp_{1,3}\bd_1 \bd_3+\bp_{1,4}\bd_1 \bd_4 \big.\\
	&\big.+\bp_{2,3}\bd_2 \bd_3+\bp_{2,4}\bd_2 \bd_4+\bp_{3,4}\bd_3 \bd_4 \big)^2 \nonumber \\
	&=\left(\bp_{1,2}\bp_{3,4}+\bp_{1,3}\bp_{2,4}+\bp_{1,4}\bp_{2,3}\right) \bd_1 \bd_2 \bd_3 \bd_4 \nonumber\\
	&\quad+(\text{terms with at least one repeated } \bd_i) \nonumber .
}
When we premultiply the last equation by $\bra{\vac}  \b_1 \b_2 \b_3 \b_4$ and postmuliply by $\ket{\vac}$, we find
\eq{
	&\bra{\vac}  \b_1 \b_2 \b_3 \b_4  \	\frac{1}{2}\left( \sum_{k<j}^4 \bp_{k,j} \bd_k \bd_j \right)^{2} \ket{\vac}\\
	&= \bp_{1,2}\bp_{3,4}+\bp_{1,3}\bp_{2,4}+\bp_{1,4}\bp_{2,3} = \haf\left(\bm{B}^{(\bm{p})}\right).
}

Let us now consider the other two cases.
For $2q=2, \ q'=2$, we have
\eq{
	\frac{\hat D^1}{1!} \frac{\hat E^2}{2!} &= \frac{1}{2}\Big(\bp_{1,2}\bd_1 \bd_2+\bp_{1,3}\bd_1 \bd_3+\bp_{1,4}\bd_1 \bd_4 \big.\\
	&\quad \big.+\bp_{2,3}\bd_2 \bd_3+\bp_{2,4}\bd_2 \bd_4+\bp_{3,4}\bd_3 \bd_4 \Big)  \nonumber \\
	&\quad \times \left(\zp_1 \bd_1+\zp_2 \bd_2+\zp_3 \bd_3+\zp_4 \bd_4 \right)^2 \nn \\
	&=\big( \bp_{1,2} \zp_3 \zp_4+\bp_{1,3} \zp_2 \zp_4+\bp_{1,4} \zp_2 \zp_3 \big. \nn \\
	&\quad \big. +\bp_{2,3}\zp_1 \zp_4+\bp_{2,4}\zp_1 \zp_3+\bp_{3,4} \zp_1 \zp_2 \big) \times \nonumber \\
	&\quad \bd_1 \bd_2 \bd_3 \bd_4 +(\text{terms with at least one repeated } \bd_i) .\nonumber
}
Likewise for $2q=0,\ q'=4$, we find
\eq{
	\frac{\hat D^0}{0!} \frac{\hat E^4}{4!}	 &= \tfrac{1}{4!} \left( \zp_1 \bd_1+\zp_2 \bd_2+\zp_3 \bd_3+\zp_4 \bd_4 \right)^4\\
	&= \zp_1 \zp_2 \zp_3 \zp_4 \bd_1 \bd_2 \bd_3 \bd_4 \nonumber 	 \\
	&\quad+(\text{terms with at least one repeated } \bd_i) .\nonumber
}
Putting together these three cases, we find that the probability amplitude for $4$ modes with one boson each is
\eq{
	\nu =& 	T	\big[ \bp_{1,2}\bp_{3,4}+\bp_{1,3}\bp_{2,4}+\bp_{1,4}\bp_{2,3}  \big.  \\
	& \quad + \bp_{1,2} \zp_3 \zp_4+\bp_{1,3} \zp_2 \zp_4+\bp_{1,4} \zp_2 \zp_3 \nonumber  \\
	& \quad  + \bp_{2,3}\zp_1 \zp_4+\bp_{2,4}\zp_1 \zp_3+\bp_{3,4} \zp_1 \zp_2 \nn \\
	& \quad +\big. \zp_1 \zp_2 \zp_3 \zp_4  \big]. \nonumber 
}
The quantity inside the square brackets is the loop hafnian of $\bar {\bm{B}}$, $\lhaf(\bar{\bm{B}})$ where 
\eq{\label{Bt}
\bar{B}_{i,j} = (1-\delta_{i,j}) B^{(\bm{p})}_{i,j}+\delta_{i,j} \zeta^{(\bm{p})}_i
}
i.e., the matrix obtained from $\bm{B}^{(\bm{p})}$ by placing the vector $\bm{\zeta}^{(\bm{p})}$ along its diagonal.

\subsubsection{Contractions: The general result}
Now let us look at an arbitrary amplitude. We want to show that the following quantity
\eq{\label{exp}
	\braket{\bm{0}|\left( \bigotimes_{j=1}^P \b_j \right) \hat \Pi_P \exp(\hat D) \exp(\hat E)|\bm{0}}	
}
is precisely $\lhaf(\bar{\bm{B}})$ where $\bar{\bm{B}}$ is defined in Eq.~(\ref{Bt}). To this end, we note that 
\eq{
	\hat \Pi_P &\exp(\hat D) \exp(\hat E) = \lhaf(\bar{\bm{B}})  \bigotimes_{j=1}^P \bd_j \\
	&\quad \quad \quad+(\text{terms with at least one repeated } \bd_i) .\nonumber
}
To show this, we argue that every possible element of the set $\text{SPM}(P)$ \emph{must} be present in the RHS of the last equation, and its coefficient must be one. We show this by construction, showing how to find an arbitrary term  $X \in \text{PMP}(P)$. Let $X$ have $q'$ loops and $q = (P-q')/2$ edges connecting different vertices. The corresponding element of the loop hafnian associated with $X$ will be found in the following polynomial 
\eq{
\frac{\hat D^q}{q!} \frac{\hat E^{q'}}{q'!}	
}
present in the Taylor expansion of Eq.~(\ref{exp}). Moreover, since the terms that are not loops need to form a matching (they cannot share any $\ad_i$) and the terms that are loops cannot be repeated, the coefficient accompanying this term has to be 1. This is readily seen by looking at the multinomial theorem in Eq.~(\ref{multinomial}), which shows that terms in which all of the $k_t=1$ have coefficient exactly equal to one.

\subsection{Summary}\label{summary}
To calculate a probability amplitude $\nu$ of $\ell$ modes, as defined in Eq.~(\ref{mudef}), we proceed as follows:
\begin{enumerate}
\item \underline{Input}: bra and ket photon numbers $\bm{m},\bm{n} \in \mathbb{Z}^\ell$, displacement vector $\bm{\alpha} \in \mathbb{\mathbb{C}^\ell}$, unitary matrices $\bm{U},\bm{U}' \in \mathbb{C}^{\ell \times \ell}$ and squeezing parameters $\bm{\lambda} \in \mathbb{R}^\ell$. For an FCFs the unitary matrices and squeezing parameters are obtained from the singular value decomposition of the matrix $\bm{A} = \left( \bm{\Omega_\out}^{1/2} \bm{O}_{D} \bm{\Omega_\inn}^{-1/2} \right)$ where $\bm{O}_{D}$ is the Duschinsky matrix and $\bm{\Omega_{\inn/\out}} $ are the matrices with the normal frequencies of the equilibrium positions of the initial/final energy surface. Finally, $\bm{\alpha} = \bm{d}/\sqrt{2}$ where $\bm{d}$ is the displacement vector between the equilibrium positions of the energy surfaces.

\item Calculate the Bloch-Messiah decomposition (implemented in, for example, \texttt{strawberryfields}\cite{killoran2018strawberry}) of the Gaussian unitary 
\eq{
\bm{Q}&= \left( \uu(\bm{U}) \otimes \id \right) \left( \s (\bm{\Lambda}) \otimes \id \right)  \left( \uu(\bm{U}') \otimes \id \right) \mathcal{\hat T}(\bm{t})\nn \\
&=\uu(\bm{\tilde U})	\s(\bm{\tilde \Lambda}) \uu(\bm{\tilde U}'),
}
to obtain $\bm{\tilde U} \in \mathbb{C}^{2\ell \times 2\ell}$, $\bm{\tilde \lambda} \in \mathbb{R}^{2\ell} $ where the parameters $t_j = \sinh^{-1} \sqrt{n_j}$. 

\item Assemble the vectors
\eq{
\bm{p} & = (m_1,m_2,\ldots, m_\ell,n_1,n_2,\ldots,n_\ell)	\in \mathbb{Z}^{2\ell},\\
\bm{\tilde \alpha} &= (\alpha_1,\alpha_2,\ldots,\alpha_\ell,\underbrace{0,0,\ldots,0}_{\ell \text{ times }}) \in \mathbb{C}^{2\ell}.
}

\item Form the following matrix and vector 
\eq{
\bm{B}&= \tu \tanh(\tl) \tu^T \in \mathbb{C}^{2\ell \times 2\ell}, \\
\bm{\zeta}&=\ta - \bm{B}\ta^* \in \mathbb{C}^{2\ell }.
}
where $\bm{\tilde{\Lambda}} = \text{diag}(\bm{\tilde{\lambda}})$.
\item Calculate the prefactors
\eq{
R &= \left( \prod_{j=1}^\ell \frac{\tanh^{n_j} t_j}{\cosh t_j} \right)^{-1},\\
T=&	\frac{\exp\left(-\tfrac{1}{2} \left\{ |\bm{\tilde{\alpha}}|^2 - \bm{\tilde{\alpha}}^\dagger \bm{B} \bm{\tilde{\alpha}}^*\right\}  \right)}{\sqrt{\prod_{j=1}^{2 \ell} \left(p_j! \cosh(\lambda_j) \right)}}.
}
\item Form the matrix $\bm{B}^{(\bm{p})} \in \mathbb{C}^{P \times P}$ and the vector $\bm{\zeta}^{(\bm{p})} \in \mathbb{C}^{P}$ according to the definition in Sec. \ref{rulem}.
\item \underline{Output}: the amplitude is given by
\eq{
\nu = R \ T \ \lhaf(\bm{\bar{B}}),	
}
where the matrix $\bm{\bar{B}} \in \mathbb{C}^{P \times P}$ is formed by taking $\bm{B}^{(\bm{p})}$ and placing the vector $\bm{\zeta}^{(\bm{p})}$ along its diagonal.
\end{enumerate}
This algorithm has been implemented as a python notebook that uses \texttt{strawberryfields} for the Bloch-Messiah decomposition and $\texttt{hafnian}$ for the loop hafnian computations\cite{fockgaussian}.

\section{Discussion}\label{sec:discussion}

The procedure described in this manuscript allows us to calculate Franck-Condon factors or, more generally, a Fock amplitude of a Gaussian unitary, of $\ell$ modes in terms of the loop hafnian of a $P \times P$ complex matrix. Using the results of Bj\"{o}rjklund \ea\cite{bjorklund2018faster}, one concludes that the complexity of calculating a particular amplitude should scale like $O(P^3 2^{P/2})$ where $P = M+N$ and $M/N$ is the total number of vibrational quanta in the excited/ground electronic energy surface. 
For zero temperature, one has $N=0$, hence the complexity is simply $O(M^3 2^{M/2})$.

The results we presented significantly improve the state-of-the art for FCFs calculations. Indeed, using the methods of Kan\cite{kan2008moments}, Huh \cite{huh2011unified} showed that an FCF can be calculated in time 
\eq{
\left(1+\Bigg\lfloor\frac{1}{2}\sum_{k=1}^\ell (n_k+m_k) \Bigg\rfloor \right) \prod_{k=1}^\ell \left[(n_k+1)(m_k+1)	\right].
}
For $n_k = m_k = 1 \ \forall k$, the above formula scales like $O(P 2^P)$; the loop hafnian methodology scales like $O(P^3 2^{P/2})$, which is significantly faster. 
For this case, one can also argue that there are very good reasons to believe one cannot improve on the $2^{P/2}$ scaling. This is because of the identity
\eq{\label{eq:bipartite}
	\lhaf \left( \left[
	\begin{array}{cc}
		\bm{0} & \bm{W} \\
		\bm{W}^T & \bm{0} \\
	\end{array}
	\right]\right) = \haf \left( \left[
	\begin{array}{cc}
		\bm{0} & \bm{W} \\
		\bm{W}^T & \bm{0} \\
	\end{array}
	\right]\right) = \text{per}(\bm{W}),
}
where per is the permanent matrix function and $\bm{W} \in \mathbb{C}^{P/2 \times P/2}$.
Based on this identity, if one could calculate the loop hafnian of an arbitrary $P \times P$ matrix in a time that scales like $\text{poly}(P) \alpha ^{P/2}$ with $\alpha<2$, one could also calculate permanents of $N \times N$ matrices in $\text{poly}(N) \alpha ^ N$ with $\alpha<2$. This would be an extremely surprising result since computer scientists have been looking at algorithms that achieve this feat for many decades (see for example Exercise 11 of Sec. 4.6.4. of the classic book by Knuth\cite{TAOCP}).

As the number of bosons \emph{per mode} is increased, using the loop hafnian algorithm from Ref. \cite{bjorklund2018faster} stops being optimal since it is designed for cases where $\bm{B}^{(\bm{p})}$ is full rank.
When the number of bosons in each mode is greater than one $p_k>1$, the matrix $\bm{B}^{(\bm{p})}$ is no longer full rank. Indeed, the matrix $\bm{B}^{(\bm{p})}$ can be of any size but its rank is at most $2\ell$. For example, if one sets $n_k = m_k = x \ \forall k$ ($P=2 \ell x$), using the algorithm from Ref. \cite{bjorklund2018faster} will require $O((\ell x)^3 2^{\ell x})$ operations versus $O(\ell (x+1)^{2 \ell+1})$ for the algorithm of Refs.\cite{kan2008moments,huh2011unified}. In this case for $x<3$ the loop hafnian algorithm from Ref. \cite{bjorklund2018faster} will outperform the algorithm from Ref. \cite{huh2011unified}

In conclusion, we provide an algorithm that is close to optimal in the low boson number per mode regime and that provides obvious computational advantages to other well known methodologies for the calculation of FCFs in the Harmonic approximation. Beyond these algorithmical results, and perhaps more importantly, our paper bridges two seemingly disparate computations. On the one hand, the calculation of Franck-Condon factors, and on the other, the calculation of the number of perfect matchings of undirected graphs with loops. The mapping developed in this manuscript should allow to import any result developed for the calculation of loop hafnians to the calculation of Franck-Condon factors and vice-versa, providing a useful link between quantum chemistry, quantum optics and computer science.
For example, using this link one can use the algorithms of Huh\cite{huh2011unified} and Kan\cite{kan2008moments} to calculate the probability amplitude of multiboson events ($p_j \gg 1$) more efficiently.  These probability amplitudes have been shown to be useful in determining graph invariants \cite{bradler2018graph}. On the flip side, and as highlighted before, in the limit of small number of bosons per mode $1 \lessapprox p_j $, one can use algorithms from graph  theory\cite{cygan2015faster,bjorklund2012counting,bjorklund2018faster} to increase the speed of Franck-Condon factor calculations.

\section*{Acknowledgements}
N.Q. thanks J.M. Arrazola, A. Bj\"orklund, A. Delgado, I. Dhand, C. Ducharme, B. Gupt, J. Izaac, N. Killoran, J. Lavoie, G.E. Moyano, M. Schuld and C. Weedbrook for fruitful discussions.

\section*{References}
\bibliographystyle{unsrt}
\bibliography{haf}
\appendix

\section{Reordering the exponentials}\label{app:reordering}

By using disentangling theorems summarized in Sec. \ref{sec:HO} we rewrite the probability amplitude in Eq. (\ref{amp1}) into a matrix element involving only destruction operators as in Eq. (\ref{amplitude}).
We follow partially the work of Kok \ea \cite{kok2001multi} generalizing their results to include also finite displacements.

We begin our calculation by acting the squeezing operator on the zero boson state using the identity in Eq.~(\ref{no1}) as follows
\eq{
	\s(\tl )\ket{\bm{0}} = 	\bigotimes_{j=1}^{2\ell}	\hat S_j(\tilde{\lambda}_j) \ket{\vac} =  \bigotimes_{j=1}^{2\ell} \tfrac{\exp\left( \tanh(\tilde{\lambda}_j) \hat a_j^{\dagger 2}/2\right)}{\sqrt{\cosh(\tilde{\lambda}_j)}} \ket{\vac},
}
where $\tilde{\Lambda}_{j,j} = \tilde{\lambda}_j$.  
Secondly, we would like to move the displacement operators to the right of the linear optics passive transformation $\uu(\tu)$. This we do by inserting a resolution of the Hilbert space identity as $\uu(\tu) \uu^\dagger(\tu) = \id $  to obtain
\sseq{
	\D(\ta)\uu&=	\bigotimes_{j=1}^{2\ell}	\hat D_j(\tilde{\alpha}_j) \uu  =\uu \uu^\dagger  \exp\left(\sum_{j=1}^{2\ell} \tilde{\alpha}_j \ad_j - \hc \right) \uu \\
	&=\uu   \exp\left(\sum_{j=1}^\ell \tilde{\alpha}_j \uu^\dagger \ad_j \uu - \hc \right) \\
	&=\uu   \exp\left(\sum_{j=1,k}^\ell \tilde{\alpha}_j \tilde{U}_{j k}^*\ad_k - \hc \right) \\
	&= \uu   \exp\left(\sum_{k}^\ell \beta_k \ad_k - \hc \right)\\
	&=\uu \bigotimes_{k=1}^\ell \hat D_k(\beta_k) = \uu \D(\bm{\beta}), 
}
where 
\eq{
	\beta_k = \sum_{j=1}^{2\ell}\tilde{U}_{j k}^* \tilde{\alpha}_j , \quad \bm{\beta} = \bm{\tilde{U}}^\dagger \ta.
}
In the last equation, and in what follows, we have omitted the argument of the passive linear operator $\uu$; if it is not specified, it is assumed to be $\tu$.

We can now reassemble our probability amplitude as
\eq{
	\nu	 = R \times \bra{\bm{p}} \uu \D(\bm{\beta}) \bigotimes_{j=1}^{2\ell}  \tfrac{\exp\left( \tanh(\tilde{\lambda}_j) \hat a_j^{\dagger 2}/2\right)}{\sqrt{\cosh(\tilde{\lambda}_j)}} \ket{\vac}.
}
Our next step is to move once more the displacement operator by writing $\id = \D^\dagger(\bm{\beta}) \D(\bm{\beta})$ next to the zero boson state and expand 
\eq{
	&\hat D_j(\beta_j) \exp\left( \tanh(\lambda_j) \hat a_j^{\dagger 2}/2\right) \ket{\vac}\\
	&= \hat D_j(\beta_j) \exp\left( \tanh(\lambda_j) \hat a_j^{\dagger 2}/2\right) \hat D_j^\dagger (\beta_j) \hat D_j (\beta_j) \ket{\vac} \nonumber\\
	&=  \exp\left( \tanh(\lambda_j) \left\{\hat D_j(\beta_j) \hat a_j^{\dagger} \hat D_j^\dagger (\beta_j) \right\}^2/2\right)  \hat D_j (\beta_j) \ket{\vac}. \nonumber 	
}
Now we use Eq.~(\ref{dispback}) to obtain $\hat D_j(\beta_j) \hat a_j^{\dagger} \hat D_j^\dagger (\beta_j) = \ad_j - \beta_j^*$ and Eq.~(\ref{no1}) to find $\hat D_j(\beta_j) \ket{\bm{0}} = e^{-|\beta_j|^2/2} e^{\beta_j \ad_j} \ket{\bm{0}} $ which allows us to write 
\eq{ 
	&\exp\left( \tanh(\tilde{\lambda}_j) \left\{\hat D_j(\beta_j) \hat a_j^{\dagger} \hat D_j^\dagger (\beta_j) \right\}^2/2\right)  \hat D_j (\beta_j) \ket{\vac} \nonumber \\ 
	&=  \exp\left( \tanh(\tilde{\lambda}_j) \tfrac{\left\{ \hat a_j^{\dagger} -\beta_j^* \right\}^2}{2}\right)  e^{-|\beta_j|^2/2}\exp\left(\beta_j \ad \right) \ket{\vac} \nonumber \\
	&=\exp\left(-\tfrac{1}{2} \left\{|\beta_j|^2 - \tanh(\tilde{\lambda}_j)\beta_j^{*2}  \right\}  \right) \\
	& \quad \times \exp\left(\tfrac{1}{2} \tanh(\tilde{\lambda}_j) \a_j^{\dagger 2}\right) \nonumber \\
	& \quad \times \exp\left(\left\{ \beta_j - \tanh(\tilde{\lambda}_j) \beta_j^* \right\} \a_j^\dagger \nonumber \right).
}

It is convenient to define 
\eq{
	\gamma_j = \beta_j - \tanh(\tilde{\lambda}_j) \beta_j^*,\quad \bm{\gamma} = \bm{\beta} - \left(\tanh \bm{\tilde \Lambda} \right) \bm{\beta}^*,
}
in order to write the amplitude as
\eq{
	\nu =& R \ T \times \bra{\bm{0}} \left( \bigotimes_{j=1}^{2\ell} \a_j^{p_j} \right) \\
	& \quad \times  \uu  \bigotimes_{j=1}^{2\ell} \exp\left(\tfrac{1}{2} \tanh(\tilde{\lambda}_j) \a_j^{\dagger 2}\right) \exp\left(\gamma_j \a_j^\dagger \nonumber \right) \ket{\vac},\\
	T=& \frac{\exp\left(-\tfrac{1}{2} \left\{ |\bm{\tilde{\alpha}}|^2 - \bm{\tilde{\alpha}}^\dagger \bm{B} \bm{\tilde{\alpha}}^*\right\}  \right)}{\sqrt{\prod_{j=1}^{2 \ell} \left(p_j! \cosh(\tilde{\lambda}_j) \right)}} ,
}
where we rewrote 
\eq{
	\bra{\bm{p}} = 	\bra{\bm{0}} \bigotimes_{j=1}^{2\ell} \frac{\a_j^{p_j}}{\sqrt{p_j!}},
}
and moved the factors of $1/\sqrt{p_j!}$ to the prefactor $T$. Also we introduced the symmetric matrix
\eq{
	\label{Bdef}
	\bm{B} = \bm{B}^T =\tu \tanh(\tl) \tu^T,
}
in terms of which we can write
\eq{
	&\sum_{j=1}^{2 \ell}|\beta_j|^2 - \tanh(\lambda_j)\beta_j^{*2} = |\bm{\beta}|^2 - \bm{\beta}^\dagger \tanh(\bm{\tilde{\Lambda}}) \bm{\beta}^*\\
	& \quad = |\bm{\tilde{\alpha}}|^2 - \bm{\tilde{\alpha}}^\dagger \left(\tu \tanh(\bm{\tilde{\Lambda}}) \tu^T \right) \bm{\tilde{\alpha}^*} =|\bm{\tilde{\alpha}}|^2 - \bm{\tilde{\alpha}}^\dagger \bm{B} \bm{\tilde{\alpha}^*} .
}
Now we move $\uu$ inside each exponential separately:
\eq{
	&\uu  \bigotimes_{j=1}^{2\ell} \exp\left(\tfrac{1}{2} \tanh(\tilde{\lambda}_j) \a_j^{\dagger 2}\right) \exp\left(\gamma_j \a_j^\dagger  \right) \ket{\vac}=\\
	&
	\uu  \bigotimes_{j=1}^{2\ell} \exp\left(\tfrac{1}{2} \tanh(\tilde{\lambda}_j) \a_j^{\dagger 2}\right) \uu^\dagger \uu  \exp\left(\gamma_j \a_j^\dagger \nonumber \right) \uu^\dagger \uu \ket{\vac }	=\\
	&    \bigotimes_{j=1}^{2\ell} \exp\left(\tfrac{1}{2} \tanh(\tilde{\lambda}_j) \uu \a_j^{\dagger 2} \uu^\dagger \right)    \exp\left(\gamma_j \uu \a_j^\dagger \uu^\dagger \nonumber \right)  \ket{\vac }	=\\
	&\exp\left(\sum_{j=1}^{2\ell} \tfrac{1}{2} \tanh(\tilde{\lambda}_j) \uu \a_j^{\dagger 2} \uu^\dagger \right)    \exp\left(\sum_{j=1}^{2\ell} \gamma_j \uu \a_j^\dagger \uu^\dagger \nonumber \right)  \ket{\vac }.
}
In the last equation, we used $\uu \ket{\vac} = \ket{\vac}$ and the identity $\uu \exp(\hat A) \uu^\dagger = \exp(\uu \hat A \uu^\dagger)$ for any $\hat A$.
For the exponential with linear argument in the destruction operators, we have 
\eq{
	\hat C &=\sum_{j=1}^{2\ell} \gamma_j \uu \a_j^\dagger \uu^\dagger = \sum_{j=1}^{2\ell} \zeta_j \ad_j ,
}
where we used the definition of the matrix $\bm{B}$ in Eq.~(\ref{Bdef}) to write 
\eq{
	\zeta _k &= \sum_{j=1}^{2 \ell} \gamma_j \tilde{U}_{kj}, \\
	\bm{\zeta} &= \tu \bm{\gamma} = \ta - \left( \tu \tanh(\tl)\tu^T \right) \ta^* = \ta - \bm{B} \ta^*.
}
For the quadratic part we have
\eq{
	\hat A&=	\frac{1}{2} \sum_{j=1}^{2 \ell}  \tanh(\tilde \lambda_j) \uu \hat a_j^{\dagger 2} \uu^\dagger = \frac{1}{2} \sum_{j=1}^{2 \ell} \tanh(\tilde \lambda_j) \left( \uu \hat a_j^{\dagger } \uu^\dagger \right)^2\nn \\
	&=\frac{1}{2} \sum_{j=1}^{2 \ell} \tanh(\tilde \lambda_j) \left( \sum_{j=1}^{2 \ell} \tilde{U}_{kj} \hat a_k^{\dagger }  \right)^2 \\
	&=\frac{1}{2} \sum_{j=1}^{2 \ell} \tanh(\tilde \lambda_j) \left( \sum_{k=1}^{2 \ell} \tilde{U}_{kj} \hat a_k^{\dagger }\right) \left(  \sum_{m=1}^{2 \ell} \tilde{U}_{mj} \hat a_m^{\dagger } \right)  \\
	&= \frac{1}{2} \sum_{k,m,j=1}^{2 \ell} \tilde{U}_{k j}  \tanh(\tilde \lambda_j)      \tilde{U}_{m j} \hat a_m^{\dagger } \hat a_k^{\dagger }\\
	&= \frac{1}{2} \sum_{k,m,j=1}^{2 \ell} \left\{ \tilde{U}_{k j }  \tanh(\tilde \lambda_j)      (\tilde{U}^T)_{jm} \right\} \hat a_m^{\dagger } \hat a_k^{\dagger }\\
	&= \frac{1}{2} \sum_{k,m}^{2 \ell} B_{kj} \hat a_m^{\dagger } \hat a_k^{\dagger },
}
where we  used once more the matrix $\bm{B}$ in Eq.~(\ref{Bdef}).

\section{Amplitudes in the coherent state basis}\label{app:coh}
Using the results from the last Appendix one can also easily calculate the following amplitude
\eq{
	\nu' =   \braket{\bm{\beta}|\mathcal{\hat D}(\bm{\alpha}) \uu(\bm{U}) \s (\bm{ \lambda}) |\bm{0}},
}
which represents an arbitrary Gaussian pure state
\eq{\label{puregaussian}
\ket{\psi_G} = \mathcal{\hat D}(\bm{\alpha}) \uu(\bm{U}) \s (\bm{ \lambda}) \ket{\bm{0}},
}
in the coherent state basis
\eq{
\ket{\bm{\beta}}	=\D(\bm{\beta}) \ket{\bm{0}}.
}
To this end we rewrite
\eq{
	\mathcal{\hat D}(\bm{\alpha}) \uu(\bm{U}) \s (\bm{ \lambda})  \ket{\bm{0}} = \frac{\exp\left(-\tfrac{1}{2} \left\{ |\bm{{\alpha}}|^2 - \bm{{\alpha}}^\dagger \bm{B} \bm{{\alpha}}^*\right\}  \right)}{\sqrt{\prod_{j=1}^{\ell} \left(\cosh({\lambda}_j) \right)}}  \times \nonumber \\
\exp(\hat A(\bm{\ad})) \exp(\hat C(\bm{\ad})) \ket{\bm{0}},	
}
where $\hat A$ and $\hat C$ are second and first degree polynomials of the destruction operators $\ad_i$ defined in the last appendix. Note that we wrote explicitly the functional dependence of the polynomials $\hat A$ and $\hat C$; this will come handy in a moment.
We now use the fact that the coherent states are left eigenstates of the destruction operators
\eq{
\bra{\bm{\beta}	}\ad_i = \bra{\bm{\beta}	} \beta_i^*
}
to write
\eq{
	\nu' =&   \braket{\bm{\beta}|\mathcal{\hat D}(\bm{\alpha}) \uu(\bm{U}) \s (\bm{ \lambda})  |\bm{0}}\\
	=& \frac{\exp\left(-\tfrac{1}{2} \left\{ |\bm{{\alpha}}|^2 - \bm{{\alpha}}^\dagger \bm{B} \bm{{\alpha}}^*\right\}  \right)}{\sqrt{\prod_{j=1}^{\ell} \cosh({\lambda}_j)}} \times \nonumber \\
	&\exp(\hat A(\bm{\beta}^{*})) \exp(\hat C(\bm{\beta}^{*})) \braket{\bm{\beta}|\bm{0}}\\	
		=&\frac{\exp\left(-\tfrac{1}{2} \left\{ |\bm{{\alpha}}|^2 - \bm{{\alpha}}^\dagger \bm{B} \bm{{\alpha}}^*\right\}  \right)}{\sqrt{\prod_{j=1}^{\ell} \cosh({\lambda}_j) }} \times \nonumber \\
		&\exp(\hat A(\bm{\beta}^*)) \exp(\hat C(\bm{\beta}^*)) \exp\left( -\tfrac{|\bm{\beta}|^2}{2}\right),
}
where we used the well known overlap between a coherent state and vacuum in the last line.

\section{The case of zero displacement: hafnians}\label{app:hamilton}
Let us consider in detail the probability amplitude 
\eq{
	\bar{\nu} &= \left. \braket{\bm{m}|\mathcal{\hat D}(\bm{\alpha} ) \uu(\bm{U}) \s (\bm{\lambda}) \uu(\bm{U}') |\bm{n} }	\right|_{\bm{\alpha} = \bm{n} = \bm{0}}\\	
	&=\braket{\bm{m}| \uu(\bm{U}) \s (\bm{\lambda})  |\bm{0} }.
}
This amplitude corresponds to a pure Gaussian state (cf. Eq. (\ref{puregaussian})) with zero displacement projected onto the Fock basis.
Using the results from Sec. \ref{sec:amplitudes} this amplitude is equal to the loop hafnian
\eq{
\bar{\nu} = T \ \lhaf(\bar{\bm{B}})
}
where we used the fact that $\bm{n}=\bm{0}$ implies $R = 1$ and also $\bm{m} = \bm{p}$. Note that since $\bm{\alpha} = \bm{0}$ then $\zeta = \bm{0}$ and thus the matrix $\bar{\bm{B}}$ defined Eq. (\ref{Bt}) has zeros in the diagonal thus we can write
\eq{
\bar{\nu} &=  \frac{\haf(\bar{\bm{B}})}{\sqrt{\prod_{j=1}^{ \ell} \left(m_j! \cosh(\lambda_j) \right)}},	\\
|\bar{\nu}|^2 &= \frac{|\haf(\bar{\bm{B}})|^2}{\prod_{j=1}^{ \ell} \left(m_j! \cosh(\lambda_j) \right)},
}
the last of which is precisely the probability calculated in Hamilton \ea \cite{hamilton2017gaussian}

\end{document}